\documentclass[useAMS,usenatbib,usedcolumn]{mnras}
\usepackage{url}
\usepackage{graphicx}
\usepackage{float}
\usepackage[normalem]{ulem}
\usepackage{xargs}
\usepackage{comment}
\usepackage{verbatim, amsmath, amsfonts, amssymb}
\usepackage[pdftex,dvipsnames]{xcolor}  % Coloured text 
\usepackage[colorinlistoftodos,prependcaption,textsize=tiny]{todonotes}
\newcommandx{\unsure}[2][1=]{\todo[linecolor=red,backgroundcolor=red!25,bordercolor=red,#1]{#2}}
\newcommandx{\change}[2][1=]{\todo[linecolor=blue,backgroundcolor=blue!25,bordercolor=blue,#1]{#2}}
\newcommandx{\info}[2][1=]{\todo[linecolor=OliveGreen,backgroundcolor=OliveGreen!25,bordercolor=OliveGreen,#1]{#2}}
\newcommandx{\improvement}[2][1=]{\todo[linecolor=Plum,backgroundcolor=Plum!25,bordercolor=Plum,#1]{#2}}
\newcommandx{\thiswillnotshow}[2][1=]{\todo[disable,#1]{#2}}
\hypersetup{
   unicode=false,                  % non-Latin characters in Acrobat book
   pdftoolbar=true,                % show Acrobat toolbar?
   pdfmenubar=true,                % show Acrobat menu?
   pdffitwindow=true,              % window fit to page when opened
   pdfstartview={FitH},            % fits the width of the page t
   pdftitle={R443130},             % title
   pdfauthor={S-PLUS},             % author
   pdfsubject={Astronomy},         % subject of the document
   pdfcreator={dvipdf},            % creator of the document
   pdfproducer={dvipdf},           % producer of the document
   pdfkeywords={S-PLUS},           % list of keywords
   pdfnewwindow=true,              % links in new window
   colorlinks=true,                % false: boxed links; true: colored
   linkcolor=red,                  % color of internal links
   citecolor=blue,                 % color of links to bibliography
   filecolor=magenta,              % color of file links
   urlcolor=cyan,                  % color of external links
   breaklinks=true,
   linktocpage
}

\title[S-PLUS: Photometric Redshifts]{Assessing the photometric redshift precision of \\ the S-PLUS survey: the Stripe-82 as a test-case.}

\author[A. Molino, M. V. Costa-Duarte, L. Sampedro et al.]{
A. Molino$^{1}$\thanks{E-mail: albertomolino.work@gmail.com}
M. V. Costa-Duarte$^{1}$, 
L. Sampedro$^{1}$, 
F. R. Herpich$^{1}$,
L. Sodr\'e Jr.$^{1}$, %\\
\newauthor{
C. Mendes de Oliveira$^{1}$, 
W.~Schoenell$^{2}$, 
C. E. Barbosa$^{1}$, 
C. Queiroz$^{3}$, 
E. V. R. Lima$^{1}$,
} 
\newauthor{
L. Azanha$^{1}$,
N. Mu\~noz-Elgueta$^{4,5}$, 
T.~Ribeiro$^{6}$, 
A.~Kanaan$^{7}$, 
J. A. Hernandez-Jimenez$^{8,1}$,
}
\newauthor{
A. Cortesi$^{9}$,
S. Akras$^{10,11}$,
R. Lopes de Oliveira$^{12,11,13}$,
S. Torres-Flores$^{4}$, 
C. Lima-Dias$^{4}$,
}
\newauthor{
J. L. Nilo Castellon$^{4}$,
G. Damke$^{4,5}$,
A. Alvarez-Candal$^{11}$, 
Y. Jim\'enez-Teja$^{11}$, 
P. Coelho$^{1}$,
}
\newauthor{
E. Pereira$^{14}$,
A. D. Montero-Dorta$^{15}$,
N. Ben\'itez$^{16}$,
T. S. Gon\c calves$^{8}$, 
L. Santana-Silva$^{8}$, 
}
\newauthor{
S. V. Werner$^{1}$,
L. A. Almeida$^{17,1}$, 
P. A. A. Lopes$^{8}$,
A. L. Chies-Santos$^{2}$,
E. Telles$^{10}$,
}
\newauthor{
Thom de Souza, R. C.$^{19}$,
D. R. Gon\c calves$^{8}$,
R. S. de Souza$^{18}$,
M. Makler$^{20}$,
}
\newauthor{
V. M. Placco$^{21,22}$,
L. M. I. Nakazono$^{1}$,
R. K. Saito$^{7}$,
R. A. Overzier$^{11,1}$,
L. R. Abramo$^{3}$}
\\
\\ Affiliations can be found after the references.}

\begin{document}
\date{Submitted to MNRAS.}

%\pagerange{\pageref{firstpage}--\pageref{lastpage}} \pubyear{2019}

\maketitle
\label{firstpage}

\begin{abstract}
In this paper we present a thorough discussion about the photometric redshift (photo-z) performance of the Southern Photometric Local Universe Survey (S-PLUS). This survey combines a 7 narrow + 5 broad passband filter system, with a typical photometric-depth of r$\sim$21 AB. For this exercise, we utilize the Data Release 1 (DR1), corresponding to 336 deg$^{2}$ from the Stripe-82 region. We rely on the \texttt{BPZ2} code to compute our estimates, using a new library of SED models, which includes additional templates for quiescent galaxies. When compared to a spectroscopic redshift control sample of $\sim$100k galaxies, we find a precision of $\sigma_{z}<$0.8\%, $<$2.0\% or $<$3.0\% for galaxies with magnitudes r$<$17, $<$19 and $<$21, respectively. A precision of 0.6\% is attained for galaxies with the highest \texttt{Odds} values. These estimates have a negligible bias and a fraction of catastrophic outliers inferior to 1\%. We identify a redshift window (i.e., 0.26$<z<$0.32) where our estimates double their precision, due to the simultaneous detection of two emission-lines in two distinct narrow-bands; representing a window  opportunity to conduct statistical studies such as luminosity functions. We forecast a total of $\sim$2M, $\sim$16M and $\sim$32M galaxies in the S-PLUS survey with a photo-z precision of $\sigma_{z}<$1.0\%, $<$2.0\% and $<$2.5\% after observing 8000 $deg^{2}$. We also derive redshift Probability Density Functions, proving their reliability encoding redshift uncertainties and their potential recovering the $n(z)$ of galaxies at $z<0.4$, with an unprecedented precision for a photometric survey in the southern hemisphere.
\end{abstract}


\section{Introduction}
\label{intro}

Modern astronomy has entered a new era of massive data acquisition. The current and new generation of redshift surveys such as the Sloan Digital Sky Survey (SDSS; \citealt{2000AJ....120.1579Y}), Pan-STARRS \citealt{2002SPIE.4836..154K}, the Dark Energy Survey (DES; \citealt{2018ApJS..235...33D}), the Large Synoptic Survey Telescope (LSST; \citealt{2008arXiv0805.2366I}), the Baryon Oscillation Spectroscopic survey (BOSS; \citealt{2009astro2010S.314S}), EUCLID \citealt{2010arXiv1001.0061R}, the Dark Energy Spectroscopic Instrument (DESI; \citealt{2013arXiv1308.0847L}) and the Javalambre-Physics of the Accelerated Universe Astronomical survey (J-PAS; \citealt{2009ApJ...691..241B}, \citealt{jpasredbook}) among others, will provide either multicolor or spectral information for millions of galaxies, enabling precise cosmological studies at different cosmic epochs. 

\vspace{0.2cm} 

In this context, photometric redshifts (photo-z) have become an essential tool in modern astronomy since they represent a quick and relatively inexpensive way of retrieving redshift estimates for a large amount of galaxies in a reasonable amount of observational time. In the last few decades, photometric redshift surveys have been mainly undertaken the following two pathways: higher wavelength resolution and moderate depth using medium-to-narrow filters versus deeper observations with poor resolution using standard broadband filters. The strong dependency between the wavelength resolution (number and type of passbands) and the achievable precision of photo-z estimates (\citealt{1994MNRAS.267..911H}, \citealt{1998ApJS..115...35H}, \citealt{2001A&A...365..681W}, \citealt{2009ApJ...692L...5B}) has inspired the design of a whole generation of medium-to-narrow multi-band photometric redshift surveys such as the Classifying Object by Medium-Band Observations-17 survey (COMBO-17; \cite{2003A&A...401...73W}), the MUltiwavelength Survey by Yale-Chile (MUSYC; \cite{Gawiser06}), the Advance Large Homogeneous Medium Band Redshift Astronomical survey (ALHAMBRA; \cite{2008AJ....136.1325M}), the Cluster Lensing and Supernovae with Hubble survey (CLASH; \cite{2012ApJS..199...25P}) and the Survey for High-z Absorption Red and Dead Sources (SHARDS; \cite{2013ApJ...762...46P}) among others, reaching photo-z estimates as accurate as $\Delta$z/(1+z)$<$0.01 for high signal-to-noise ratio (S/N) galaxies. If we take into account that the new generation of multi narrow-band photometric surveys will surpass the photometric-depth of current spectroscopic redshift surveys such as SDSS (r$<$18, \cite{2015ApJS..219...12A}) and will provide increasingly more accurate photometric redshift estimations, it is expected that the current picture of the physical processes governing the assembling history and evolution of the Universe that we have today will be soon re-built. Meanwhile, very deep broadband photometric observations such as the Hubble Deep Field (HDF; \cite{1995AAS...187.0901F}), the Hubble Ultra-Deep Field (HUDF; \cite{2006AJ....132.1729B}), the Cosmic Assembly Near-infrared Deep Extragalactic Legacy Survey (CANDELS; \cite{2011ApJS..197...35G}), the Hubble Extreme Deep Field (XDF; \cite{2013ApJS..209....6I}), the Hubble Frontiers Field program (HFF; \cite{2017ApJ...837...97L}) or the REionization LensIng Cluster Survey (RELICS; \citealt{Coe2019}) among others, even with a limited photo-z accuracy of $\Delta$z/(1+z)$>$0.05, have extended our current knowledge on the formation and evolution of galaxies all the way back to a $z\sim$10-12.

\vspace{0.2cm} 

The Southern-Photometric Local Universe Survey (S-PLUS\footnote{\url{http://www.splus.iag.usp.br}}; \textcolor{blue}{Mendes de Oliveria et al., (subm.)}) is an on-going observational program aiming at imaging 9300 deg$^{2}$ from the Southern Hemisphere, with a 0.8 m and 1.4 x 1.4 $deg^{2}$ Field-of-View robotic telescope (hereafter T80S). The S-PLUS is equipped with an optical filter system (Figure \ref{splusfilters}) designed to perform accurate photometric stellar spectral-type classifications (\cite{2004A&A...419..385B}, \cite{2006MNRAS.367..290J}, \cite{MarinFranch12}, \cite{Gruel12}). The system covers the entire optical range, from 3700$\AA$ to 9000$\AA$, with a total of 12 photometric bands. In particular, the system includes 5 standard broad-bands (u, g, r, i \& z) useful to constrain the spectral continuum of the sources and 7 narrow ($\sim$150\AA-width) bands (J0378, J0395, J0410, J0430, J0515, J0660, J0861) to trace the $[OIII]$, Ca H+k, H$\delta$, G-band, Mgb Triplet, H$\alpha$ and Ca Triplet features, respectively. Although the filter system was originally designed for the star classification, its wavelength resolution renders possible to attain accurate distance estimates for nearby and low-redshift galaxies and for galaxies at specific redshift windows (see Section \ref{photozmain} of this paper for a thorough discussion). Filter transmission curves for the S-PLUS survey can be accessed through its website\footnote{\url{http://www.splus.iag.usp.br/en/camera-and-filters/}}. Along with this filter system, T80S is equipped with an optical imager composed of a 9k x 9k pixel-array and a 0.55”/pixel scale (see \cite{MarinFranch15}, \cite{2019A&A...622A.176C}, \textcolor{blue}{Mendes de Oliveira et al., (subm.)}, for more details). The combination of all these three elements (e.g., accurate photo-z estimates for nearby galaxies, a wide Field-of-View and a detector with a large pixel-array) makes S-PLUS a powerful dataset to carry out systematic IFU-like analysis for all spatially resolved galaxies, eliminating biases due to preselected samples. 

\vspace{0.2cm}

The S-PLUS project will simultaneously conduct several sub-surveys, in most cases using different observational strategies, aiming at tackling different scientific cases. Here we briefly summarize the main characteristic and scientific goals of the \texttt{Main Survey} (hereafter \texttt{MS}) and we refer the interested reader to \textcolor{blue}{Mendes de Oliveira et al., (subm.)} for an in-depth description of these observational programs. The \texttt{MS} is motivated to conduct extragalactic science. The superb photometric redshift estimations provided by the S-PLUS survey in the nearby Universe (see Section \ref{photozmain}) will allow to complement current analysis of the large-scale structure of the Universe, without relying on bright colour-based pre-selected spectroscopic galaxy samples. 
The footprint has been designed to have large  overlapping areas with already existing or forthcoming deep extra-galactic surveys such as DES, KiDS, ATLAS and LSST. These common regions will serve both for calibration purposes of the S-PLUS observations and to provide improved photometric redshifts for objects in these fields down to i=21 magnitudes. 
%%DES \citep{2017arXiv171206209T}, KiDS \citep{2015A&A...582A..62D}, ATLAS \citep{2015MNRAS.451.4238S} and LSST \citep{2008arXiv0805.2366I}.

\vspace{0.2cm}

The superb precision of the S-PLUS photometric redshift in the nearby Universe, compared to similar 4-5 passband photometric redshift surveys of similar depth (see \citealt{Molino18} for a detailed discussion), will make possible to revisit fundamental aspects of extragalactic astronomy, such as the formation of the structures in the Universe at (i.e., groups or clusters of galaxies) and the large-scale structure of the nearby Universe at the present epoch, due to the large observational surveyed area (i.e., 8000 deg$^{2}$). The data and results presented in this work correspond to those obtained from the Data Release I (DR1) of the \texttt{MS}. This means that the precision achieved by the S-PLUS photometric redshifts might not necessarily correspond to the one achievable using the data from another sub-survey, where the number of filters and/or the photometric-depth of the observations may vary substantially. 

\vspace{0.2cm}

This paper is organized as follows. Section \textcolor{blue}{2} opens this paper with a description of the dataset utilized in this exercise. In Section \textcolor{blue}{3}, we introduce the photometric redshift code chosen for this work, the new updates that were required to fulfill the specifications of the S-PLUS survey, the evaluation matrices and the discussion of the precision achieved. Along with this, this section also includes a comparison with other catalogues, a discussion about the role of narrow-band filters and the identification of a redshift window where our estimates get boosted due to a simultaneous detection of emission lines. Section \textcolor{blue}{4} motivates the computation of Probability Density Functions and presents a number of statistical tests to prove its reliability. Section \textcolor{blue}{5} explains how the absolute magnitude and the stellar mass content is calculated along with a quantification about how uncertainties in redshift propagate into these estimates. Section \textcolor{blue}{6} is devoted to the characterization of the quality of input the multiband photometric data. This exercise includes a check-up of the photometric zero-points and validation of the photometric uncertainties. Final Sections, i.e., \textcolor{blue}{7}, \textcolor{blue}{8}, \textcolor{blue}{9} include, respectively, a list of directions about how to use and access the data along with a number of tables detailing the performance of our photometric redshift estimates.

\vspace{0.2cm}

Unless specified otherwise, all magnitudes here are presented in the AB system. Throughout this work, we have adopted the cosmological model provided by the \cite{2014A&A...571A..16P} with parameters ($h_{0}$, $\Omega_{M}$, $\Omega_{\Lambda}$, $\Omega_{K}$) = (0.70, 0.31, 0.69, 0.00).

\section{Data}
\label{data}

Through the following subsections, we describe the data utilized in this work. Initially, in Section \ref{Observations}, we describe our observations in terms of area and depth. Then, in Section \ref{photometry}, we briefly review both the S-PLUS image-reduction and calibration pipeline and the main aspects of our photometric pipeline. Section \ref{S82surveys} summarizes the available public data in the Stripe-82 from other surveys or facilities. Finally, in Section \ref{controlsample}, we describe the spectroscopic redshift galaxy sample selected for the characterization of our photometric redshift estimations.         

\subsection{Observations}
\label{Observations}

We choose the Stripe-82 region to characterize the expected precision of the S-PLUS photometric redshifts. The area, which is a 2.5 degree wide and 270 degree long stripe along the Celestial Equator in the Southern Galactic Cap (i.e., $-50^{o}<\alpha<59^{o}$, $-1.25^{o}<\delta<1.25^{o}$), has been extensively observed by a large number of facilities (see Section \ref{S82surveys}), counting with abundant public spectroscopic redshift information for a large number of galaxies, down to a magnitude $r$=22. Therefore, it is ideal for data verification purposes. 

\vspace{0.2cm}

\begin{figure*}
\includegraphics[width=18.0cm]{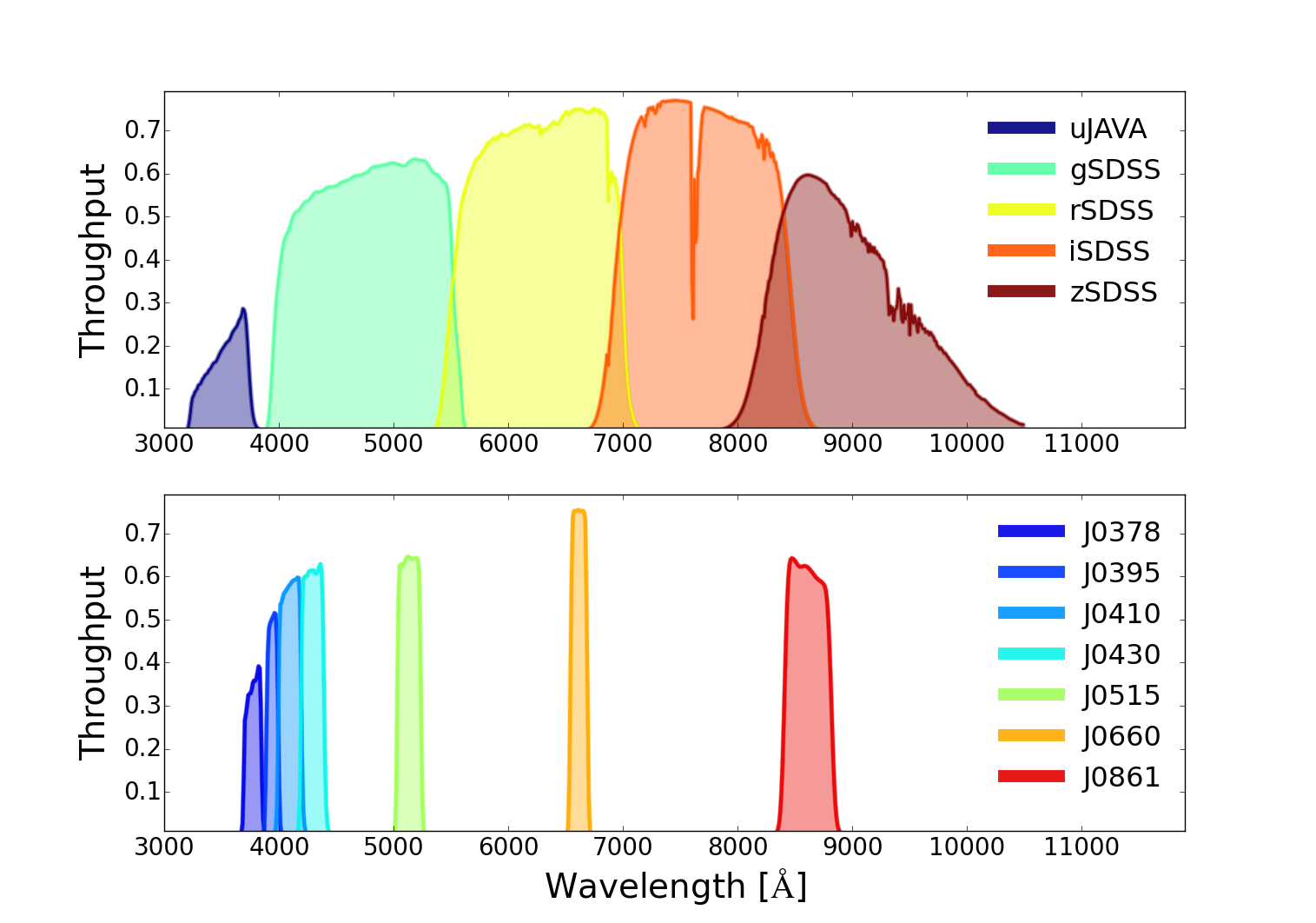}
\caption[S-PLUS filter system.]{The S-PLUS filter system. Top and bottom panels show, respectively, the 5 broad- and 7 narrow-band filters.}
\label{splusfilters}
\end{figure*}

In this work, we make use of the S-PLUS Data Release 1 (DR1), which corresponds to a total area of 336 $deg^{2}$ across the Stripe-82 region, divided in a total of 170 individual and contiguous pointings. The observations were gathered during two periods: from August to November, 2016 and from August to October 2018, as part of the \texttt{Main Survey (MS)} (see \textcolor{blue}{Mendes de Oliveira et al. (subm.)} for further information). Every pointing is observed with our 12-band filter system (see Figure \ref{splusfilters}). As described in \textcolor{blue}{Sampedro et al., (in prep.)}, the \texttt{MS} observations reach a typical photometric-depth of $r\sim$21 magnitudes in the 5 broad-bands and $r\sim$20.5 magnitudes in the 7 narrow-bands, for sources detected with a significance larger than S/N$>$3. Since these observations also correspond to S-PLUS verification data, it is expected a certain level of inhomogeneity in the data in terms of depth, due to an unequal amount of co-added images. Although the Stripe-82 regions do not contain very bright stars, several of the DR1 images do present saturated stars due to the integration time requested for the \texttt{MS}. Since a masking of saturated stars was not applied to our images, photometry for detections near saturated stars might be compromised. Although accurate photometry for bright stars (i.e., $r<12$) is performed in the \texttt{Ultra-Short survey (USS)}, we do not make use of that information, restricting the data to the specification of the \texttt{MS}.

\subsection{Photometric Catalogues.}
\label{photometry}

In this section we revise the procedure adopted in S-PLUS to produce the photometric catalogues used in this work. We refer the interested reader to \cite{2019A&A...622A.176C} or to \textcolor{blue}{Mendes de Oliveria et al., (subm.)} for a further discussion on the data-reduction process, and to \textcolor{blue}{Sampedro et al., (in prep.)} for a thorough discussion on the photometric calibration and photometry extraction. 

\vspace{0.2cm}

In short, raw individual images are reduced on a daily-basis following a standard procedure (i.e., bias, flat-field and fringing subtraction), where cold and hot pixels, cosmic rays and satellite tracks are detected and masked out. Final PSF-homogenized and astrometriced co-added science images are generated as a combination of individual exposures from each filter using, respectively, \texttt{SExtractor} \citep{2010asclsoft10064B} and \texttt{PSFEx} \citep{2013ascl.soft01001B}, \texttt{SCAMP} \citep{2010ascl.soft10063B} and \texttt{SWARP} \citep{2010ascl.soft10068B} software.    

\vspace{0.2cm}

Photometric calibration is carried out in a two-step process. A relative calibration is first computed using a new technique presented in \textcolor{blue}{Sampedro et al. (in prep.)}, specifically developed for wide-field multi-band photometric surveys such as S-PLUS, J-PLUS \citep{2019A&A...622A.176C}, J-PAS \citep{jpasredbook} and LSST \citep{2008arXiv0805.2366I}. This technique utilizes already available multi-band photometric catalogues from other surveys, such as SDSS \citep[see][]{2007AJ....134..973I, 2008ApJ...674.1217P}, Pan-STARRS \citep{2012ApJ...756..158S}, DES \citep[see][]{2018ApJS..235...33D, 2018AJ....155...41B} and KiDS \citep{2015A&A...582A..62D}, to predict the colours of main sequence stars in the S-PLUS filter system. Finally, a homogeneous absolute calibration is performed bringing the S-PLUS photometry to that from Gaia \citep{2017A&A...599A..50A}. As demonstrated in \textcolor{blue}{Sampedro et al., (in prep.)}, this technique is capable to provide zero-point estimates with uncertainties around 2-3\%, without needing to rely on long observational campaigns to observed standard stars in every filter and pointing.  

\vspace{0.2cm}

The S-PLUS photometric pipeline produces aperture-matched PSF-corrected multi-band photometric catalogues for every field. This pipeline is similar to those presented in other surveys such as the ALHAMBRA survey \citep{Molino14}, the CLASH survey \citep{Molino17} or the J-PLUS survey \citep{Molino18}. This photometric pipeline, which is based on the \texttt{SExtractor} software, provides different photometries defined in various types of apertures to accommodate a large number of scientific cases. Sources identified on detection images, combination of the reddest (griz) broad-band images. Photometric uncertainties, initially derived by \texttt{SExtractor} software, are re-computed on an image-by-image basis, to account for the correlation among adjacent pixels introduced during the image-reduction and co-adding process. These re-estimations serve to both to improve posterior SED-fitting analysis (i.e., setting realistic uncertainties) and to provide more reliable photometric upper-limits. In this paper, we have relied on the most restricted apertures derived in S-PLUS (i.e., \texttt{auto\_restricted}) to compute photometric redshifts, since these apertures provide accurate colours and high signal-to-noise magnitudes \citep{Molino17}.

\subsection{The Stripe-82: archival data.}
\label{S82surveys}

The Stripe-82 is a 2.5 degree wide and 270 degree long stripe along the Celestial Equator in the Southern Galactic Cap (i.e., $-50^{o}<\alpha<59^{o}$, $-1.25^{o}<\delta<1.25^{o}$), extensively observed by a large number of facilities, covering a wide range in wavelength. In the following sections, we summarize the different photometric and spectroscopic programs which have performed observations over this area, providing basic information and references.

\subsubsection{Imaging data.}
\label{photosample}

As part of the Data Release 7 (SDSS/DR7; \citealt{2009ApJS..182..543A}) and as part of the SDSS/Supernovae Survey \citep{2008AJ....135..338F}, the Sloan Digital Sky Survey has repeatedly scanned the Stripe-82 region, reaching an imaging depth two magnitudes deeper than the main SDSS survey (\citealt{2014ApJ...794..120A}; \citealt{2014ApJS..213...12J}; \citealt{2016MNRAS.456.1359F}). In addition to the SDSS imaging, this region has been observed in the optical wavelength range by other programs such as the CFTH/MegaCam Stripe-82 Survey (CS82; \citealt{2013MNRAS.433.2545E}) and the Dark Energy Survey (DES; \citealt{2017arXiv171206209T}). At other wavelengths, the Stripe-82 has been covered by GALEX in the far- and near-UV \citep{2007ApJS..173..682M}, by the United Kingdom Infrared Deep Sky Survey (UKIDSS; \citealt{2007MNRAS.379.1599L}), the VISTA/CFHT Stripe 82 Near-infrared Survey \citep{2017ApJS..231....7G} and the UKIDSS Deep eXtragalactic Survey (DXS; \citealt{2007MNRAS.379.1599L}) in the NIR, by the Wide-field Infrared Survey Explorer (WISE; \citealt{2010AJ....140.1868W}), the Spitzer HETDEX Exploratory Large-area Survey (SHELA, \citealt{2016ApJS..224...28P}) and the Spitzer IRAC Equatorial Survey (SpIES, \citealt{2016ApJS..225....1T}) in the MIR. At longer wavelengths, in the FIR, the Herschel Stripe 82 Survey (HerS; \citealt{2014ApJS..210...22V}), and the HerMES Large Mode Survey (HeLMS; \citealt{2016MNRAS.462.1989A}). In the microwaves, the Atacama Cosmology Telescope (ACT; \citealt{2010ApJ...722.1148F}) and the Very Large Array (VLA; \citealt{2011AJ....142....3H}; \citealt{2016MNRAS.460.4433H}). In Radio wavelengths, the Caltech-NRAO Stripe 82 Survey (CNSS; \citealt{2016ApJ...818..105M}). Finally, in the X-rays domain, the Stripe-82 has also been covered by Chandra and XMM-Newton  (\citealt{2014MNRAS.439.1212F}; \citealt{2016ApJ...817..172L}).

\subsubsection{Spectroscopic redshifts.}
\label{specsample}

Along with the imaging data presented in the previous section, the Stripe-82 already has a large number of spectra, with tens of thousands of redshift measurements from SDSS \citep{2018ApJS..235...42A}, 2SLAQ \citep{2005MNRAS.360..839R}, 2dF \citep{2001MNRAS.328.1039C}, 6dF \citep{2004MNRAS.355..747J}, DEEP2 \citep{2013ApJS..208....5N}, VVDS \citep{2005A&A...439..845L}, and PRIMUS \citep{2011ApJ...741....8C}, SDSS-III Baryon Oscillation Spectroscopic Survey (BOSS; \citealt{2013AJ....145...10D}), SDSS-IV/eBOSS \citep{2017ApJS..233...25A} and WiggleZ \citep{2010MNRAS.401.1429D}. Further details about the photometric properties (i.e., colours and depths) of these spectroscopic redshift samples can be found in Section 3.2 of \cite{2018MNRAS.475.3613S}. 

\subsection{The spectroscopic control sample.}
\label{controlsample}

In order to characterize the precision and reliability of our photometric redshift estimations, we have compiled a sample of $\sim$100k galaxies with known spectroscopic redshifts. As explained in Section \ref{photozmain}, in this work we are using the \texttt{BPZ2} code with a library of regular galaxy models (i.e., neither AGN nor QSO models included) to compute our photometric redshifts\footnote{The precision for QSO and AGNs will be estimated in two separate papers: \textcolor{blue}{Queiroz et al., (in prep.)} and \textcolor{blue}{Nakazono et al., (in prep.).}}. In order to decontaminate the original compilation of galaxies from AGNs and QSOs, we made the following exercise. First, we ran the \texttt{BPZ2} code using the ONLY\_TYPE = YES mode, to redshift all galaxy models in its library to the corresponding spectroscopic redshift value. Based on the S-PLUS photometry for each galaxy, the \texttt{BPZ2} code searches for the model that minimizes the differences between data and models. Finally, all sources with very poor fitting (i.e, very high $\chi^{2}$ values) are discarded since these sources may correspond either to AGN or QSOs, poor-photometry sources, mismatched sources, variable sources or galaxies with extreme colours. It is worth stressing that although this selection may exclude some real galaxies from the analysis, it serves to eliminate potential contaminants that are not the main goal of this paper which is to characterize the real performance of the S-PLUS survey for regular galaxies.

\vspace{0.2cm}

As stressed in \textcolor{blue}{Mendes de Oliveria et al., (subm.))} or in \cite{2019A&A...622A.176C}, one of the main advantages of using photometric redshift surveys, such as S-PLUS and J-PLUS, instead of spectroscopic redshift samples in the study of the nearby universe, is that the former ones present a larger completeness and homogeneity in their samples since they do not depend on any previously selected sample. In terms of the characterization of the expected performance of a photometric redshift survey, this effect needs to be considered since the availability of complete spectroscopic surveys down to faint magnitudes is very scarce. Therefore, results from deep photometric redshift surveys for the faintest magnitude bins have to be interpreted with care since spectroscopic samples (used for the validation) tend to be dominated by selection effects. Fortunately, for the specific case of this work, this situation has a marginal impact. One reason for choosing the Stripe-82 for the validation of our photometric redshifts, was that this region has spectroscopic redshift information for a large fraction of relatively faint galaxies (i.e., r$>$20). This means that the S-PLUS observations are similar in terms of depth and colour-coverage to those provided by the compiled spectroscopic sample.  

\vspace{0.2cm}

\begin{figure*}
\includegraphics[width=7.4cm]{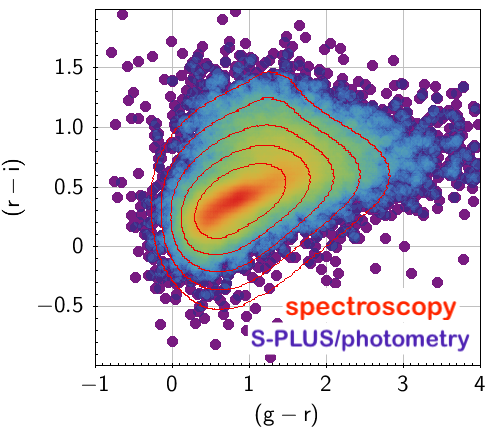}
\includegraphics[width=7.45cm]{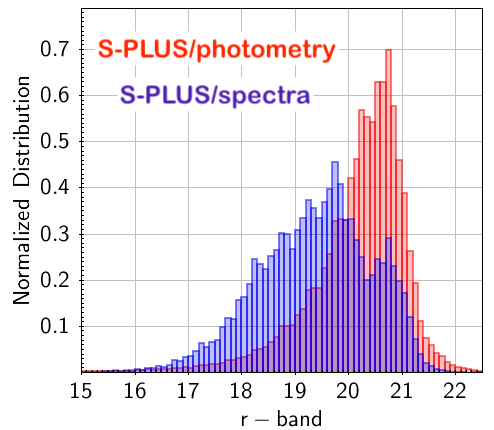}
\includegraphics[width=7.45cm]{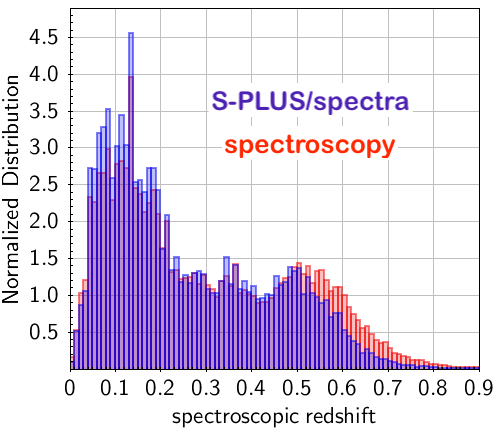}
\includegraphics[width=7.65cm]{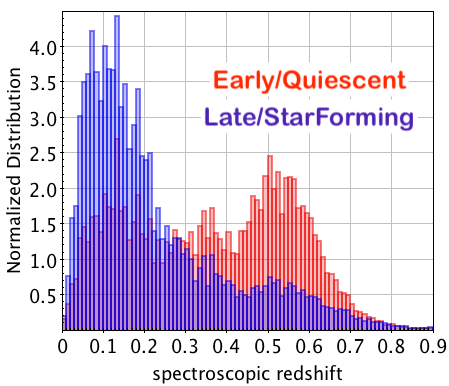}
\caption[]{Validation of the spectroscopic redshift sample. Top-left: comparison of the colour-colour distribution for both the spectroscopic redshift (contours) and the photometric (dots) samples. Top-right: Magnitude distribution of galaxies in the spectroscopic (blue) and photometric (red) samples. Bottom-left: Redshift distribution of both samples. Red histogram corresponds to all galaxies with spectroscopic redshift information within the S-PLUS footprint. Blue histogram represents the fraction of those galaxies detected in the S-PLUS observations. Bottom-right: Spectral-type distribution of galaxies as a function of redshift, with early/quiescent galaxies in red and late/starforming galaxies in blue.}
\label{speczsample}
\end{figure*}

To prove this statement, we have compared the colour, magnitude, redshift and spectral-type distribution of both photometric and spectroscopic samples. On the top left panel of Figure \ref{speczsample}, we have represented a colour-colour distribution (i.e., $g$-$r$ versus $r$-$i$) of both samples. In order to facilitate its visualization, the number density of sources from the spectroscopic sample (a.k.a., ``spectroscopy'') represented with red contours whereas the photometric sample (a.k.a., ``S-PLUS/photometry'') has been colour-coded using circular markers. As seen from this panel, both samples present a very similar distribution in this colour-colour space. Interestingly, for the reddest objects in the photometric sample (i.e., $g-r>3$), there seems to be very few galaxies with spectroscopic information. We notice that galaxies with such extreme colours may not be well represented by the results presented through this section. 

\vspace{0.2cm}

Top right panel of Figure \ref{speczsample} shows the normalized magnitude distribution for the samples introduced before. The photometric depth of the S-PLUS survey (a.k.a., ``S-PLUS/photometry'') is represented by the red histogram, whereas that of the spectroscopic sample (a.k.a., ``S-PLUS/spectra'') is represented by the blue histogram. In order to be self-consistent when comparing magnitudes among datasets (i.e., due to differences in filters and photometric apertures), we have decided to use the S-PLUS magnitudes to define the photometric-depth of both samples. Therefore, the blue histogram represents the S-PLUS magnitudes for those galaxies detected in the S-PLUS survey with a spectroscopic redshift measurement. It is worth mentioning that this comparison serves to understand whether the spectroscopic sample utilized in this work (i.e., to characterize the expected photo-z precision) can be considered representative of the entire S-PLUS survey. As expected, whereas the photometric sample (i.e., red histogram) shows a smooth and single-peak distribution at a around $r=$20.5-21.0 magnitudes, the spectroscopic sample is multi-modal and shallower, with a main peak at a magnitude $r\sim$19.5 and a secondary (less pronounced) peak at a magnitude $r\sim$20.5. This comparison can be interpreted as follows: Although it is true the majority of galaxies in the spectroscopic sample is clustered in brighter magnitude bins (i.e., 18$<r<$20), this may still have a large number of galaxies with magnitudes close to the S-PLUS limiting magnitude (i.e., rightmost tail reaching magnitude $r=$22.). In other words, the spectroscopic sample shows a high density of galaxies at those intermediate magnitudes where the S-PLUS survey can still detect galaxies with a high signal-to-noise and, a progressively decreasing sample of galaxies with spectroscopic information in the close to the survey detection limit where the quality and completeness of our photometry could be compromised. Finally, it is worth mentioning that this heterogeneity in the data does not represent an issue for the scope of this paper\footnote{We are adopting an SED-fitting approach to compute photo-z.}, but might affect other redshift estimates based on learning process where resembling samples are needed for training purposes at all magnitude and redshift bins.

\vspace{0.2cm}

The bottom left panel of Figure \ref{speczsample} shows the redshift distribution for both samples. As before, the blue histogram (a.k.a., ``S-PLUS/spectra'') corresponds to the subsample of galaxies in the spectroscopic redshift sample detected by the S-PLUS observations whereas the red histogram (a.k.a., ``spectroscopy'') to the whole spectroscopic sample within the S-PLUS footprint. This comparison serves to understand the selection effect in redshift space (i.e., the completeness) imposed by the photometric depth of the S-PLUS observations. As clearly seen in this panel, while most galaxies at $z<0.5$ are detected in the S-PLUS images, there is a large fraction of galaxies missing at $z>0.5$. This is a clear sign of a selection function in redshift for the S-PLUS data. While this issue will be fully address in Section \ref{completABz}, here we can conclude the following. The spectroscopic sample utilized in this work seems to be sufficient for the scope of this paper because, unlike most surveys, here it is the observations that limit the depth of the spectroscopic sample and not the other way around. In other words, we might be able to characterize the photo-z precision for those $z>0.5$ galaxies typically detected in S-PLUS to the photometric depth of its observations.

\vspace{0.2cm}

Finally, the bottom right panel of Figure \ref{speczsample} shows the spectral-type distribution of galaxies as a function of redshift. As before, these samples correspond solely to the galaxies in the spectroscopic sample detected in the S-PLUS observations. In order to facilitate its visualization, the red histogram corresponds to those galaxies classified by the \texttt{BPZ2} code as early/quiescent types and the blue histograms to those classified as late/star-forming types. In order to be able to know the spectral-type of each galaxy in the spectroscopic sample, we did a similar exercise as that presented at the beginning of this section. We ran the \texttt{BPZ2} code using the ONLY\_TYPE = YES mode, to redshift all galaxy models in its library to the corresponding spectroscopic redshift value. Based on the S-PLUS photometry for each galaxy, the \texttt{BPZ2} code estimated the most likely model. We finally labeled as ``early'' to those galaxies with a spectral-type \texttt{Tb}$\leq$6.5 and as ``late'' to those with a spectral-type \texttt{Tb}$>$6.5. The results from this figure can be interpreted as follows: As expected, early-type galaxies are more smoothly distributed over the entire redshift range, presenting two prominent peaks at $z=$0.15 \& $z=$0.55; being the last one associated with the Luminous Red Galaxies (i.e., LRG). Beyond that, the S-PLUS selection function makes the distribution to decline rapidly, with a very limited sample beyond $z=$1.0. On the other hand, we observe that intermediate- and late-type galaxies are more clustered at $z<$0.25, with a steadily shrinking sample at $z\geq$0.25. Although it is true that at $z\geq$0.5 the spectroscopic sample utilized in this work will be dominated by red/early-type galaxies, the number of blue/late-type galaxies will still be large enough to guarantee a robust characterization of our photo-z at those redshift ranges. We refer the reader to Sec. \ref{completABz} for a further discussion on this topic.   

\vspace{0.2cm}

In the light of what has been presented and discussed above, we consider that the spectroscopic sample utilized in this paper can be considered as representative (i.e., colour, magnitude, redshift and spectral-type terms) for the universe observed by the S-PLUS survey. Therefore, the characterization of the photo-z performance we present in the following sections, can be considered as a good estimate of the survey.  

\section{Photometric Redshifts}
\label{photozmain}

A photometric redshift (hereafter photo-z) represents an indirect estimate of the distance to an extragalactic source based on a discrete assemble of fluxes (or magnitudes) measured for an astronomical object when it is observed through a particular filter system. Along with the distance (or redshift), this technique typically also provides an estimate of the spectral energy distribution (i.e., SED) of the source in consideration. The combination of these two pieces of information (e.g., redshift and SED) makes this technique ideal for galaxy evolution studies. From its first application in the 1960s \citep{1962IAUS...15..390B}, this technique has experienced a relatively long history. Nowadays, photo-z have become an essential tool in modern astronomy since they represent a quick and almost inexpensive way of retrieving redshift estimates and SED identification for a large amount of extragalactic sources in a relatively small amount of observational time.

Although it is true that redshift estimations from galaxy colours are more uncertain than those obtained directly from a spectrum, this situation has been gradually improved. Multi medium- or narrow-band surveys such as COMBO-17 \citep{2001A&A...365..681W}, COSMOS-21 \citep{2007ApJS..172....9T}, ALHAMBRA \citep{2008AJ....136.1325M}, COSMOS-30 \citep{2009ApJ...690.1236I}, S-PLUS (\textcolor{blue}{Mendes de Oliveria et al., (subm.)}) or J-PLUS \citep{2019A&A...622A.176C}, can now achieve statistical photo-z uncertainties of $\Delta$z/(1+z)=0.01 for high S/N galaxies. The new generation of multi-band surveys, such as J-PAS \citep{jpasredbook} and PAU \citep{2014MNRAS.442...92M}, which utilizes a photometric systems composed by $\sim$60 100$\AA$-width narrow-band filters, can provide ``very low resolution spectra" achieving statistical photo-z uncertainties as low as $\Delta$z/(1+z)=0.003 for millions of galaxies, down to a magnitude $r$$\sim 22.5$ (\citealt{2016MNRAS.456.4291A}, \citealt{2018arXiv180904375E}).

\vspace{0.2cm}

As discussed in \cite{2012SPIE.8451E..34Z} and in \cite{2014MNRAS.442.3380C}, current photo-z techniques can be broadly divided in two main categories: SED-fitting and training-based algorithms. On the one side, template fitting approaches such as \texttt{BPZ} \citep{2000ApJ...536..571B}, \texttt{EAZY} \citep{2008ApJ...686.1503B}, \texttt{GAZELLE} \citep{2009MNRAS.396..462K}, \texttt{GOODZ} \citep{2010ApJ...724..425D}, \texttt{Hyperz} \citep{2000A&A...363..476B}, \texttt{Le Phare} (\citealt{2002MNRAS.329..355A}; \citealt{2006A&A...457..841I}), \texttt{LRT} (\citealt{2008ApJ...676..286A}, \citealt{2010ApJ...713..970A}), \texttt{ZEBRA} \citep{2006MNRAS.372..565F}, \texttt{IMPZ} \citep{2004MNRAS.353..654B} and \texttt{CZR} (\citealt{2001AJ....122.1151R}; \citealt{2004ApJS..155..243W}), estimate photo-z by finding the best match between the observed and the predicted magnitudes (or colours) of galaxies according to a library of galaxy models (or SED templates). The main advantage of these template-based codes is that they can be applied without needing large and high-quality spectroscopic training samples. However, inaccurate estimations of the filter transmission curves or faulty libraries of templates may severely affect the performance of these techniques. 

On the other side, machine learning methods such as \texttt{PR} (\citealt{1995AJ....110.2655C}; \citealt{2005ApJS..158..161H}), \texttt{NN/kNN} \citep{2008ApJ...683...12B}, \texttt{KR} (\citealt{2007MNRAS.382.1601W}; \citealt{2009MNRAS.397..520W}), \texttt{ArborZ} \citep{2010ApJ...715..823G}, \texttt{GPs} (\citealt{2009ApJ...706..623W}; \citealt{2010MNRAS.405..987B}), \texttt{MS} \citep{2009ApJ...695..747B}, \texttt{ANNs} (\citealt{2003MNRAS.339.1195F}; \citealt{2004PASP..116..345C}), \texttt{MLP} \citep{2004A&A...423..761V}, \texttt{SVMs} \citep{2005PASP..117...79W}, \texttt{WGE} \citep{2011MNRAS.418.2165L}, \texttt{SCA} \citep{2009MNRAS.398.2012F},  \texttt{TPZ} \citep{2013MNRAS.432.1483C}, \texttt{SOMz} \citep{2014MNRAS.438.3409C}, among others, have the advantage of being easier to include extra information (apart from magnitudes and colours), such as galaxy profiles or concentrations in the computation of redshifts. However, these methodologies are only reliable within the limits of the training dataset, making uncertain its extrapolation to different magnitude, redshift or wavelength ranges. Therefore, they are highly disadvised for surveys or datasets with small training samples.  

\vspace{0.2cm}

As emphasized before, this paper aims at describing the usability of the S-PLUS photometric redshifts in extragalactic studies. Therefore, rather than formating this work as another photo-z-challenge paper (where the performance of several codes are presented), we have preferred to apply solely a single well-tested and well-known photometric redshift code, keeping the focus on the data themselves rather than on the specific systematics each codes may display. A discussion about the benefits of combining several photo-z codes for the S-PLUS data and the optimal way of combining such information will be addressed in a separate paper.

\vspace{0.2cm}

Through the following sections, we provide a description of the potential of the S-PLUS multi-band photometric data for SED and redshift estimates of galaxies in the nearby Universe and we discuss the role it can play in extragalactic astronomy. We start by describing in \ref{bpzcode} and \ref{nearuniv} the code we have used to compute photometric redshifts and several updates necessary to adequate the software to the needs of S-PLUS. In \ref{metric} we described the metric adopted to characterize the performance of our estimates. Section \ref{testing} presents a throughout description of the results achieved as a function of a number of variables, such as magnitude, redshift, spectral-type and \texttt{Odds}. Section \ref{photozdepth} we calculate the photo-z depth of S-PLUS, providing a forecast of the number of expected galaxies in the survey with a given photo-z precision. Section \ref{photozcomparison} is devoted to compare our results with those from other previous works on the Stripe-82 using similar datasets. In Section \ref{narrowbands}, we quantify the improvement in our estimates due to the increase in the wavelength resolution provided by the 7 narrow-band filters. Finally, Section \ref{zWindows} highlights the possibility of using specific redshift windows where the photo-z precision gets improved due to the detection of emission-lines from galaxies. 

\subsection{The \texttt{BPZ2} code.}
\label{bpzcode}

We rely on the Bayesian Photometric Redshift (\texttt{BPZ2}) code (\cite{2000ApJ...536..571B}, \cite{2006AJ....132..926C}) to compute our photo-z estimates. \texttt{BPZ2} is a Bayesian template-fitting code where a likelihood function coming from the comparison between data ($D$) and models ($T$) is weighted by an empirical luminosity-based prior, as indicated in Eq. \ref{bpzequation}: 

\begin{equation} 
p(z|D,m_{0}) \propto p(z,T|m_{0}) \times p(D|z,T) 
\label{bpzequation}
\end{equation}
\vspace{0.1cm}

where $p(z|D,m_{0})$ represents the full posterior distribution (or PDF), $p(z,T|m_{0})$ the likelihood, $p(D|z,T)$ the prior and $m_{0}$ the apparent magnitude of the galaxy. As discussed below, these PDFs surpass traditional point-like estimates, enhancing the reliability of statistical analysis based on photo-z. The characterization of these distribution functions will be addressed in Section \ref{metric} and Section \ref{PDFs}.

\vspace{0.2cm}

In this work, we use the (\texttt{BPZ2}) code which has already been applied to other astronomical surveys (e.g., ALHAMBRA \citep{Molino14}, J-PAS \citep{2016MNRAS.456.4291A}, CLASH \citep{Molino17} and J-PLUS \citep{Molino18}) showing excellent results. Compared to its public version\footnote{\url{http://www.stsci.edu/~dcoe/BPZ/}}, \texttt{BPZ2} includes the following updates: it is computationally faster and its photo-z estimates are more robust. It includes a new library of galaxy templates composed by 5 early- and 9 late-type models (see Section \ref{nearuniv}), including emission lines and dust extinction. The opacity of the intergalactic medium is applied as described in \cite{1995ApJ...441...18M}. In addition, it provides an estimate of both the absolute magnitude and the stellar mass content of galaxies based on the most likely redshift and spectral-type solution. PDFs are now stored using a Hierarchical Data Format (HDF5), which is more efficient than the previous ASCII files. It also includes new priors derived from several datasets and initially applied to the ALHAMBRA survey. We refer the reader to \cite{Molino14} for more details about \texttt{BPZ2}.

\vspace{0.2cm}

As discussed by several authors (e.g., \cite{2009ApJ...691..241B}, \cite{2010MNRAS.406..881B}, \cite{2014MNRAS.437.3490M}, \cite{2018arXiv180901669T}, among others), high-precision cosmological studies based on photometric redshifts require these estimates to be robust; as much in terms of precision (i.e., small $\sigma_{z}$) as in terms of accuracy (i.e., $\mu_{z}\sim$0), and a limited fraction of catastrophic outliers. As demonstrated in several works, the \texttt{Odds} parameter from the \texttt{BPZ} code serves precisely to that purpose (\cite{2009ApJ...691..241B}, \cite{Molino14}, \cite{2015MNRAS.453.1136J}, \cite{2015MNRAS.453.2515A}). As defined in \cite{2000ApJ...536..571B}, the \texttt{Odds} of a galaxy corresponds to the ratio between the integrated probability within a redshift interval (i.e., $\Delta_{z}$) around the most probable value (i.e., $z_{p}$) in the probability distribution function (i.e., $p(z)$), over the entire probability distribution. This expression is presented as follows:

\begin{equation}
Odds = {{\int_{z_{p}-\Delta_{z}}^{z_{p}+\Delta_{z}}} \,\,p(z) \,dz \over {\int_{z_{1}}^{z_{2}}} \,\,p(z) \,dz},
\label{oddsdefinition}
\end{equation} 
\vspace{0.1cm}

where $z_{1}$ and $z_{2}$ correspond to the minimum and maximum redshift values, respectively, considered in the analysis. Based on equation \ref{oddsdefinition}, narrow distributions will result in values close to 1 since much of their integrated probabilities will be contained in the redshift interval $\Delta_{z}$. Oppositely, very broad or multi-modal distributions will result in values close to 0 since the fraction of their integrated probability will be small. Therefore, the \texttt{Odds} can be understood as a quality parameter where, the closer to 1, the more reliable (i.e, the less uncertain) the photo-z determination is. 

\vspace{0.2cm}
 
Finally, \texttt{BPZ2} allows the user to fine-tune the redshift interval over which to integrate the \texttt{Odds}, calibrating these estimates to the characteristic of any dataset. The integration interval has typically been defined as twice the expected precision of the photo-z estimates: $\Delta_{z}$ = 2$\times \sigma_{z}$. In its previous version, \texttt{BPZ2} used a fixed $\Delta_{z}$=0.06 interval, since that was the typical precision of photo-z estimates at the time (\cite{1997hsth.conf..175S}, \cite{1999ApJ...513...34F}, \cite{2003AJ....125..580C}, \cite{2006AJ....132..926C}, among others). With the tremendous improvement in the precision of these photo-z estimates from surveys including many medium and/or narrow-passbands, the definition of this integration interval had to be updated. In this work, we adopted an interval $\Delta_{z}$=0.02, since this is the averaged expected precision for most galaxies in the nearby Universe (see Section \ref{testing}).

\subsection{Updates for the nearby Universe.}
\label{nearuniv}

The \texttt{BPZ2} code has been applied to a large number of datasets, from intermediate (\cite{2014A&A...562A..86J}, \cite{Seoane17}) to high redshift (\cite{2012Natur.489..406Z}, \cite{2014ApJ...793L..12Z}, \cite{2015MNRAS.453.1136J}). Although \texttt{BPZ2} has always excelled as one of the most robust photo-z codes (e.g., \cite{2010A&A...523A..31H}), it has been reported by several authors (internal communications) that it under performs in the nearby Universe. Whereas it performs well in terms of precision (i.e., small $\sigma_{z}$), it may under-perform in terms of accuracy (i.e., $\mu_{z}$ $\neq$0). In other words, \texttt{BPZ2} may successfully identify galaxies at the same redshift (e.g., in a cluster) but assigning to them a biased redshift. As concluded in \cite{Molino17}, from a systematic study of galaxies in massive clusters using \texttt{BPZ2}, this effect can be explained as a consequence of an incomplete library of SED models. In the absence of proper models, \texttt{BPZ2} may compensate the differences in colour between models and observations by redshifting or blueshifting the templates, until reaching a mathematical minimization.   

\vspace{0.2cm}

While characterizing the performance of our photometric redshifts, we noticed a similar effect. After binning the galaxies in magnitude and/or in redshift, we discovered that most early-type galaxies retrieved a rather large bias in the error distribution. A deeper inspection showed that the redshift-colour space for early-type galaxies was not properly covered by the previous templates. In the light of this event, we decided to incorporate additional models in our library and test its new performance. Fortunately, we noticed these models increased the accuracy of our estimates, completely eliminating the previous bias (i.e, $\mu_{z}$ $<$0.1\%) at all magnitude and redshift ranges. 

\vspace{0.2cm}

\begin{figure}
\includegraphics[width=8.0cm]{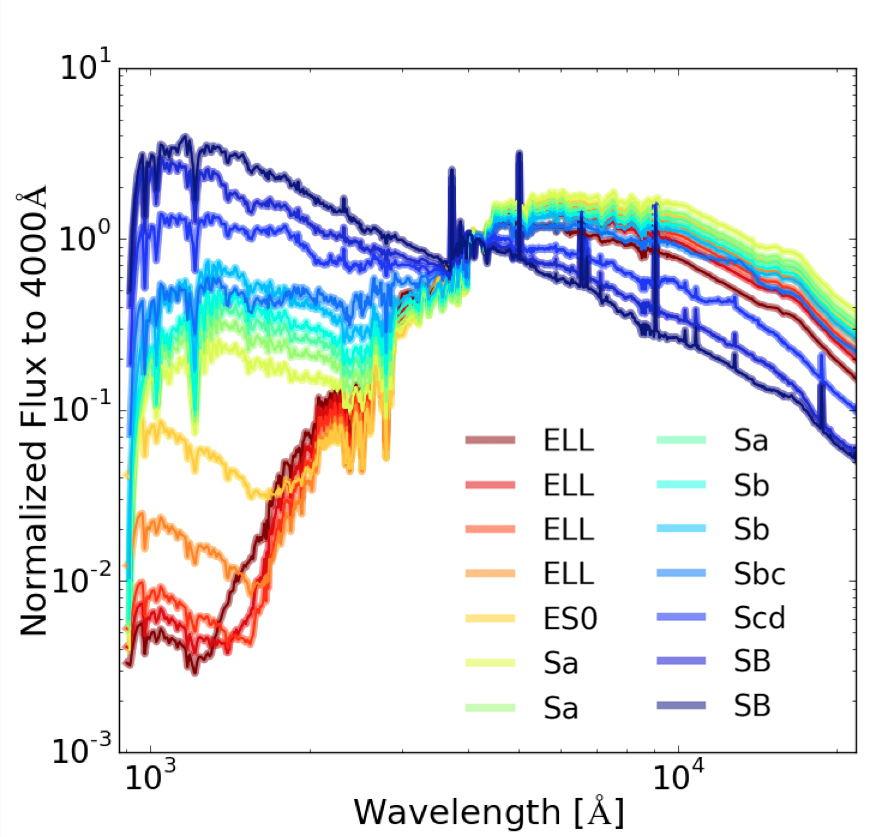}
\caption[]{Library of SED models utilized in this work. In order to improve the colour coverage of low-to-intermediate redshift early-type galaxies, 4 additional templates were incorporated to the previous library of \texttt{BPZ2}. Models include emission lines and dust extinction.}
\label{newSEDs}
\end{figure}

Additionally, we realized that the precision (i.e., $\sigma_{z}$) obtained for the early-type models was poorer than that obtained for the late-type ones. Due to the depth of our images, we noticed the signal was limited (if any) at the bluest wavelengths for the most red galaxies in our fields. Effect that could limit the performance of our photo-z estimates due to the fact that we may be relying on a subset of filters for its computation. In order to disentangling the effect of a limited signal from the representativeness of these models among red galaxies, we make use of the ALHAMBRA-Gold catalog\footnote{\url{http://cosmo.iaa.es/content/ALHAMBRA-Gold-catalog}}. This catalog includes a sample of $\sim$1000 early-type galaxies observed with a 23-band optical filter system to a depth of $r$ $\sim$25 magnitudes. By selecting galaxies down to a magnitude similar to that of the S-PLUS (i.e., $r<$22), we assure all galaxies in ALHAMBRA have a high signal-to-noise photometry in all the red and blue filters. Based on the ALHAMBRA photometry \citep{Molino14}, we ran \texttt{BPZ2} using this new library of SED models. This new library improved the previous precision, presented in \cite{Molino14}, by a factor of 2.5 for galaxies with a magnitude $r<$ 17, and a factor of 1.5 for galaxies with magnitudes in between 17$<r<$21. Additionally, it reduced, up to an order of magnitude, the bias for galaxies at low-redshift ($\mu_{z}\sim$0.00 for $z<$0.3). These results served to proved that the limited performance of our new red templates was solely due to limited photometric-depth of the S-PLUS observations. The new library of SED models, utilized in this work, is shown in Figure \ref{newSEDs}.  

\vspace{0.2cm}

Finally, the previous \texttt{BPZ2} prior was extended to include galaxies with magnitudes brighter than $r<$18. So, adequating it to the needs of the new local universe multi-band photometric redshift surveys such as S-PLUS and J-PLUS.  

\vspace{0.2cm}

\subsection{Evaluation Metric.}
\label{metric} 

Although there are several recent works where elaborated metrics are defined for the characterization of photometric redshift estimates (e.g., \cite{2014MNRAS.442.3380C} and \cite{2014MNRAS.445.1482S}), in this work we prefer to adopt a simpler, more intuitive metric as that presented in \cite{Molino14}. We address the characterization of our photometric redshifts in two independent steps. Whereas in Section \ref{testing} we treat them as single-point estimates (i.e., based on the most likely redshift solution), in Section \ref{PDFs} we analyze its performance treating them as probability distribution functions (PDFs). 

\vspace{0.2cm}

For the description of photometric redshift as single-point estimates, we rely on the Normalized Median Absolute Deviation (NMAD; see Eq. \ref{nmadeq}), which represents a robust measurement of the accuracy reached by a set of estimates \citep{2008ApJ...686.1503B}. It is worth stressing that a typical photometric redshift error distribution has extended tails, clearly departing from a pure Gaussian distribution, in addition to a relatively large fraction of outliers. The NMAD estimator manages to get a stable estimate of the spread of the core of the photo-z distribution without being affected by catastrophic errors. The accuracy of this estimator (i.e., $\sigma_{NMAD}$) is defined as:
 
%\begin{equation}
%\sigma_{NMAD} = 1.48 \times \text{median} \left(\frac{|\delta z - %\text{median}(\delta z)|}{1+z_\text{s}} \right)
%\label{nmadeq}
%\end{equation} 

\begin{equation}
\sigma_{NMAD} = 1.48 \times median({\left | \delta z - median(\delta z) \right | \over 1+z_{s}})
\label{nmadeq}
\end{equation}

\vspace{0.2cm} 

being $\delta z$ = $z_{b}$ - $z_{s}$, $z_{b}$ the Bayesian photometric redshift and $z_{s}$ the spectroscopic redshift. \footnote{For the sake of keeping the notation simple, we will adopt through this paper \textit{$\sigma$} or \textit{$\sigma_{z}$} when referring to the $\sigma_{NMAD}$}. Along with this, it is also important to quantify its precision ($\mu$) to identify any systematic bias in the redshift estimations. Finally, we describe the expected fraction of catastrophic errors ($\eta$) which, in this work, is defined as: 

\begin{equation}
\eta = {\left | \delta z \right | \over 1+z_{s}} > 5 \times \sigma_{NMAD} 
\label{outlier1def}
\end{equation}

\vspace{0.2cm} 

Finally, for the description of our photometric redshifts as probability distribution functions, we make use, in Section \ref{HPD}, of the Highest Probability Density (HPD) to measure the reliability of our PDFs encoding real redshift uncertainties.

\subsection{Photo-z Performance.}
\label{testing}

Through the following subsections, we present a number of tests describing the performance of the S-PLUS photometric redshifts based on the observations described in Section \ref{data}. These analyses describe the observed performance as a function of the $r$-band magnitude, the redshift and the \texttt{Odds} parameter from the \texttt{BPZ2} code. In all three cases, we will first estimate the average precision for all types of galaxies, and then splitting the sample in different spectral-types (i.e., early/red and late/blue types)\footnote{The spectral-type classification is done according to the most likely template selected by the \texttt{BPZ2} code using the \textit{ONLY\_TYPE=yes} mode}. Later on, in subsection \ref{photoz3D}, we will further elaborate these analyses showing the expected performance of our photo-z in a multi-dimensional space, through the combination of all the aforementioned variables. These diagrams will serve to identify sub-regions where the photo-z performance gets improved due to the given wavelength-resolution of the filter system (see Section \ref{zWindows} for a further discussion). Finally, in Section \ref{photozdepth}, and based on the \texttt{Odds} parameter, we will forecast the photometric-redshift-depth of the survey, i.e., predicting the total amount of galaxies expected in the S-PLUS survey with a minimum (maximum) photometric redshift precision as a function of the magnitude and/or redshift.   

\subsubsection{As a function of the magnitude and redshift.}
\label{photozR}

We study the dependence of the photo-z precision as a function of the apparent $r$-band magnitude and redshift. Firstly, this analysis serves to understand how our photo-z estimates become affected by the photometric noise in our data. Secondly, it also reflects the importance of the (inhomogeneous) wavelength resolution of our filter system sampling the SED of sources, since we expect to detect galaxies with similar apparent magnitudes but with different redshifts. For this exercise we define several magnitude bins ranging from 14.5 $<r<$ 21.5, and redshift values in between 0$<$z$<$1. These limits are chosen with the purpose of avoiding both very bright and too faint galaxies whose photometry could be compromised. 

\vspace{0.2cm}
  
\begin{figure*}
\includegraphics[width=17.0cm]{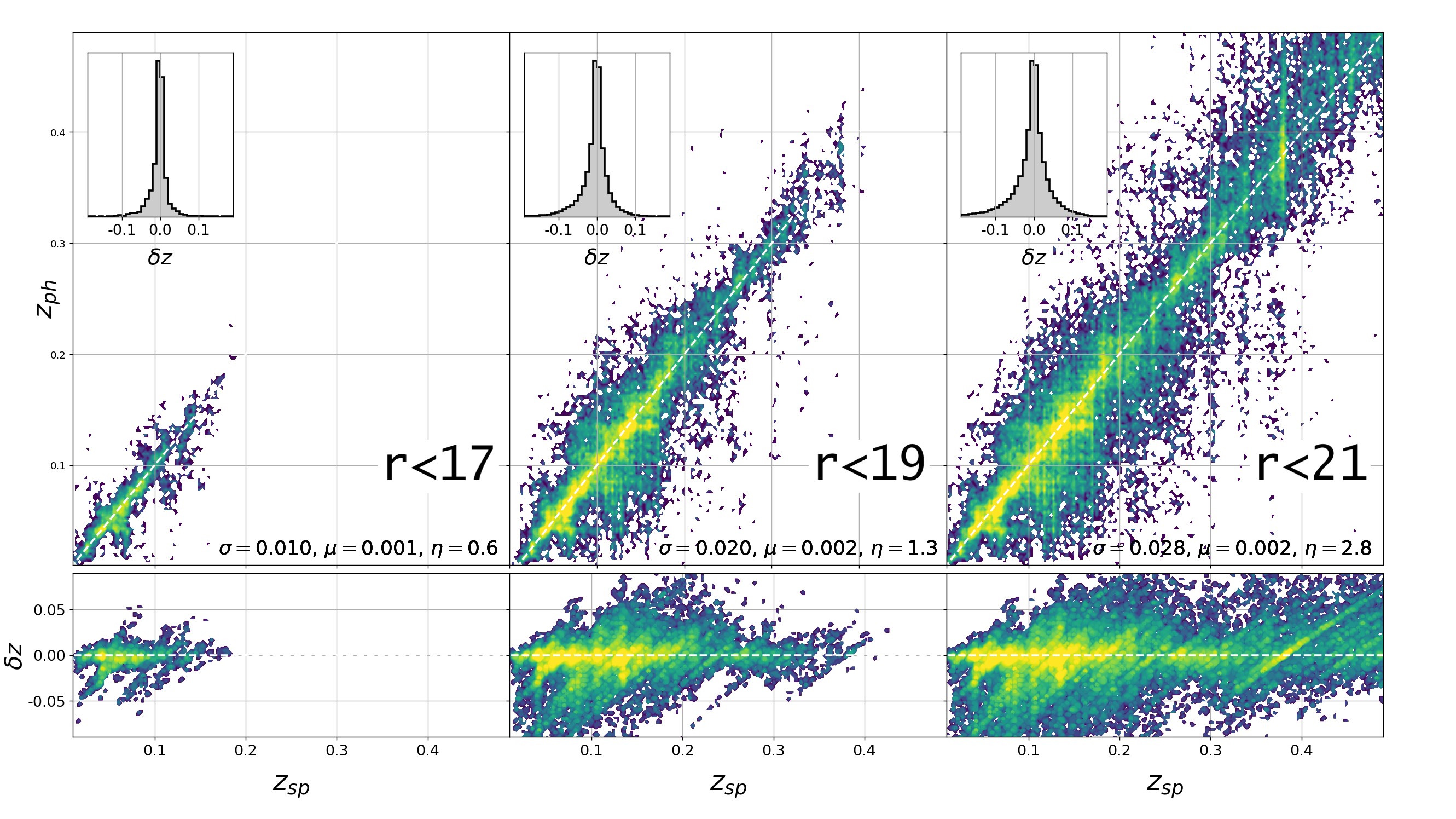}
\caption[]{Photometric Redshift Precision of the S-PLUS survey as a function of the r-band magnitude, for three different intervals. Colours correspond to a logarithmic number density of sources. As indicated in the inner labels, the expected dispersion for galaxies with magnitudes brighter than $r<17$, $r<19$ \& $r<21$ is $\delta_{z}/(1+z)<0.01$, $\delta_{z}/(1+z)<0.02$ \& $\delta_{z}/(1+z)<0.03$, respectively. Top left inner panel shows the error distribution. The symmetry of these distributions indicate that the total accumulated bias is always $<1\%$.}
\label{photozprecisionABz}
\end{figure*}

 Although a full description of the results extracted from this exercise can be found in Table \ref{phzacctable0}, Table \ref{phzacctable1} \& Table \ref{phzacctable2}, here we extract a few interesting results. Averaging over all types of galaxies and redshifts, we find a precision of $\sigma_{z}$ $\sim$1\%, $\sigma_{z}$ $\sim$2\% and $\sigma_{z}$ $\sim$3\% for galaxies with apparent magnitudes $r<$17, $r<$19 and $r<$21, respectively. This behaviour is expected since it reflects that the lower the signal-to-noise of the detection is, the more uncertain becomes its redshift estimation. In addition, it is observed a negligible bias ($\mu_{z}\sim$0.1\%) as a function of the magnitude, which indicates that the photo-z estimates are very accurate (see discussion in Section \ref{bpzcode}). Interestingly, for the brightest magnitude bins ($r<16$), we find that early-type galaxies reach a superb precision of $\sigma_{z}\sim$0.6\%; reinforcing the role of S-PLUS for clustering detecting in the nearby Universe. In Figure \ref{photozprecisionABz} we show the performance of our photo-z estimates for three different magnitude intervals ($r<$17, $r<$19 and $r<$21). Similarly, we find a precision of $\sigma_{z}\sim$1.5\% and $\sigma_{z}\sim$3\% for galaxies with a redshift $z<$0.05 \& $z<$0.5, respectively. Interestingly, we observe that early-type galaxies with redshifts below $z$$<$0.1 or $z$$<$0.5 reach a precision of $\sigma_{z}$ $\sim$1.0\% or $\sigma_{z}$ $\sim$2.0\%, respectively. Finally, it is worth mentioning that the obtained fraction of catastrophic outliers (defined as Eq.\ref{outlier1def}) was always smaller than a few percents, with a clear dependence with the magnitude and the redshift.

\subsubsection{As a function of the \texttt{Odds}.}
\label{oddsphotoz} 

\begin{figure}
\includegraphics[width=8.0cm]{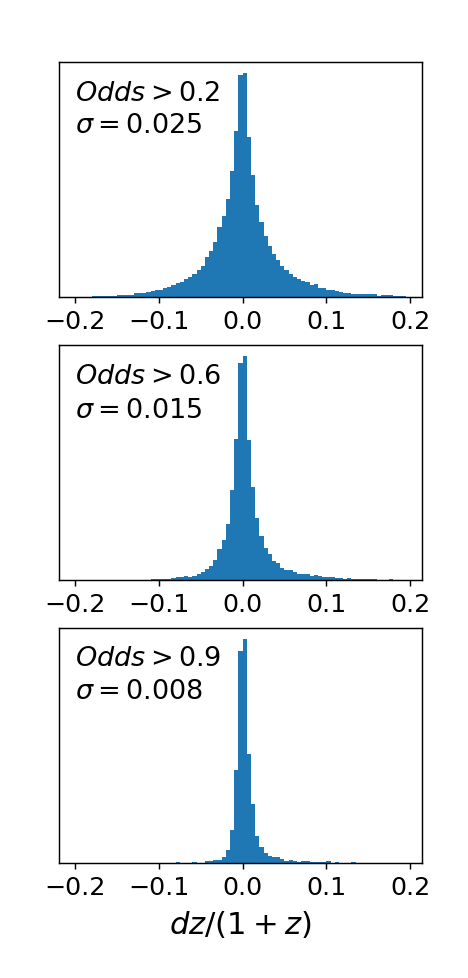}
\caption[dz]{Photo-z error distribution function for samples with different \texttt{Odds} cuts. As indicated in the legend, the higher the \texttt{Odds} value, the narrower the distribution or, in other words, the more accurate the photo-z predictions. A precision of $\sigma_{z}$=0.8\%, 1.5\% and 2.5\% is found for galaxies with values \texttt{Odds}$>$0.9, \texttt{Odds}$>$0.6 and \texttt{Odds}$>$0.2, respectively.}
\label{sigmaz}
\end{figure}

As discussed in Section \ref{bpzcode}, the \texttt{Odds} parameter renders possible the selection of clean samples of galaxies with precise (small $\sigma_{z}$) and accurate (small $\mu_{z}$) photo-z estimates. We analyze the performance of our photo-z estimates as a function of this parameter, as much globally as for different spectral-types. In Figure \ref{sigmaz}, we present the resulting photo-z error distribution function for samples with different \texttt{Odds} cuts. As indicated in the legend, the higher the \texttt{Odds} value is, the narrower the distribution is or, in other words, the more accurate the photo-z predictions are. A precision of $\sigma_{z}$=0.8\%, 1.5\% and 2.5\% is found for galaxies with \texttt{Odds}$>$0.9, \texttt{Odds}$>$0.6 and \texttt{Odds}$>$0.2, respectively. As in the previous section, we find a very small fraction of catastrophic outliers always smaller than a few percents. Additionally, Tables \ref{phzacctable0} \& \ref{phzacctable3} describe in detail the observed performance as a function of the \texttt{Odds}. 

\subsubsection{As a function of the magnitude, redshift and \texttt{Odds}.}
\label{photoz3D}

Finally, in this section we analyze the performance of our photo-z estimates combining different variables: $r$-band magnitudes, redshift and the \texttt{Odds} parameter. To motivate this exercise, it is worth mentioning that, by selecting specific bins in the magnitude-redshift space, it turns out possible to better understand the limitation in the photo-z estimates given by the photometric-depth or the filter-system of a survey. For example, it is expected that within a given magnitude bin there will exist galaxies at different redshifts. By selecting galaxies with a specific magnitude (i.e., signal-to-noise), it is feasible to isolate the contribution to the photometric redshift uncertainties coming from the wavelength resolution given by our filter system. Similarly, by selecting galaxies at the same redshift interval, we can isolate the impact of the sources detected with different signal-to-noise in the photo-z estimates.  
 
\vspace{0.2cm}

\begin{figure*}
\includegraphics[width=17.cm]{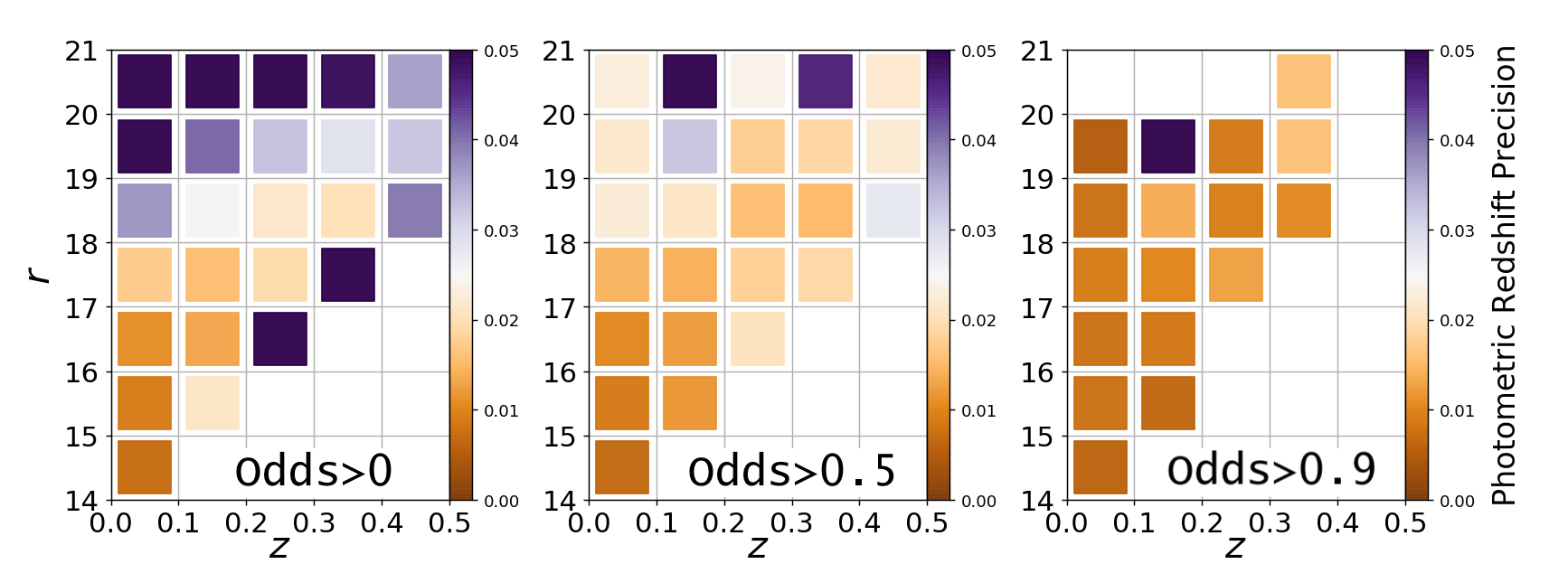}
\caption[]{Photo-z precision for different Odds cuts}
\label{sigmazOdds}
\end{figure*}

In Figure \ref{sigmazOdds} we represent the observed photo-z precision as a function of the $r$-band magnitude, redshift and the \texttt{Odds} parameter. From left to right, a selection criteria of \texttt{Odds}$>0.0$, \texttt{Odds}$>0.5$ and \texttt{Odds}$>0.9$ has been imposed. The photo-z precision is colour-coded as indicated by the vertical colour-bar. As expected, on the one hand, an overall improvement at all magnitude and redshift bins is observed as galaxies with a higher \texttt{Odds} value are selected. Therefore, it becomes feasible to retrieve samples of galaxies with a given maximum photo-z error at specific magnitude-redshift windows. On the other hand, in every individual subsample, the average precision decreases as we move upwards or rightwards, since galaxies in those bins will be progressively fainter. Interestingly, however, there are bins in which the precision abruptly improves (or worsen). These fluctuations reflect the inhomogeneous detectability of spectral features by our filter system. An example of these windows is presented in Section \ref{zWindows}. These diagrams make possible to understand which the expected precision for the S-PLUS photo-z estimates is within specific magnitude-redshift bins.  

\subsubsection{Spectral-type misclassification.}
\label{confusmat}

In this section we investigate how the uncertainties in the redshift estimations may affect the spectral-type classification of sources or, in other words, how uncertainties in the redshift space may translate into uncertainties in the spectral-type space. This is an important piece of information when deriving spectral-type-dependent statistical analysis, since it represents the capacity of distinguishing among models, acting like a spectral-type resolution indicator.

\vspace{0.2cm}

\begin{figure*}
\includegraphics[width=17.5cm]{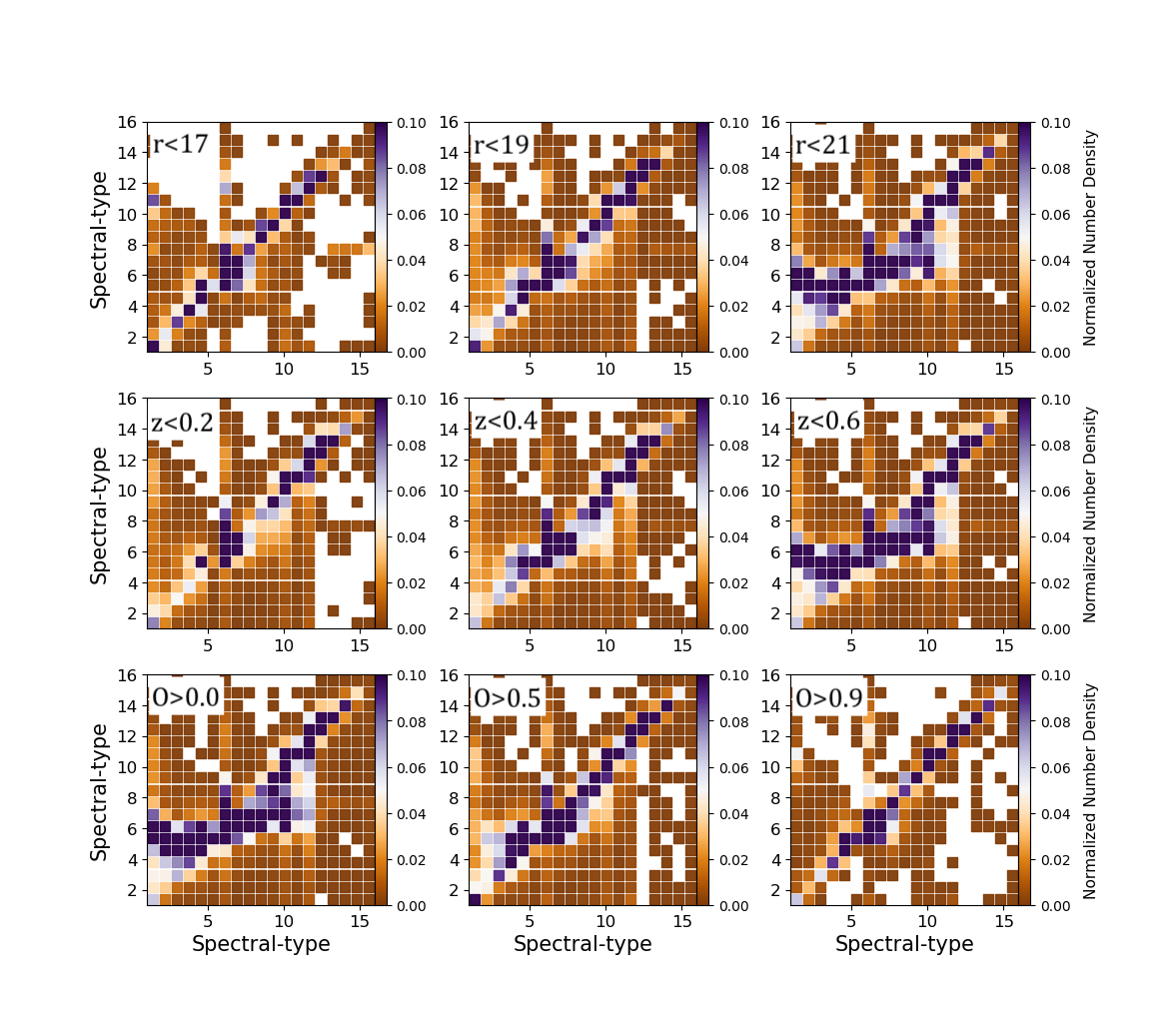}
\caption[]{Spectral-type misclassification due to uncertainties in redshift space. Horizontal axes correspond to the classification based on spectroscopic redshifts and vertical axes to those based on photometric redshifts.}
\label{confussionmat}
\end{figure*}

In order to cope with this goal, we used once more the spectroscopic sample presented in Section \ref{specsample}, running the \texttt{BPZ2} code twice on it. First, using the \texttt{ONLY\_TYPE = YES} mode redshifting all SED models to the exact redshift value. Then, we use its normal mode allowing \texttt{BPZ2} to predict the most likely redshift for each galaxy according to the S-PLUS data. Since all other configuration parameters but this are kept the same during both runs, the so-observed variations in the spectral-type classification come from the uncertainties in redshift space. Trying to make this analysis more useful, we present the results as a function of the $r$-band magnitude ($r$), the spectroscopic redshift ($z$) and the \texttt{Odds} parameter ($O$). Likewise, we split each one of these groups in three subsamples ($r<17, 19, 21$, $z<0.2, 0.4, 0.6$ and $O>0.0, 0.5, 0.9$), to see how the precision classifying the spectral-type of sources evolves with these variables. Finally, in order to facilitate the visualization of these results, as illustrated in Figure  \ref{confussionmat}, we rely on simple confusion matrices where the initial (using spectroscopic redshifts)  and final (using photometric redshifts) classifications are displayed, respectively, horizontal and vertically. The normalized number density of sources is colour-coded in each panel.

\vspace{0.2cm}

We observe that galaxies with bright magnitudes (top left), at low redshift (intermediate left) and with high $\texttt{Odds}$ values (bottom right) preserve its original classification, since most of these galaxies fall in matrix diagonal. This is an expected result since that galaxy population typically has a high signal-to-noise photometry and many of the most important spectral-features (e.g., D4000) are still mapped by the filter system. These results progressively worsen as we move to fainter, higher in redshift and lower in $\texttt{Odds}$ value galaxies. More galaxies populate non-diagonal positions in these confusion matrices, indicating the presence of misclassifications due to degeneracies. This is also expected as the photometric information available to constrain the redshift of galaxies becomes scarcer (i.e., early-type galaxies become to be non-detected in the bluest filters) or more uncertain (i.e., larger photometric noise), making more unfeasible to properly identify the real SED of source. In the worst case scenario presented here (i.e., galaxies with magnitude $r<21$, $z<0.6$ and $\texttt{Odds}>0.0$) where the overall degeneracy among models is larger than in the previous cases, we highlight an interesting finding. Whereas early-type galaxies suffer a large degeneracy among them, late-type galaxy tend to preserve its original classification. This issue might be explained by the limited depth of the bluest filters, where faint red galaxies are typically non-detected due to the D4000-break while blue galaxies are still detected since their SEDs are more luminous at those wavelengths.       
\subsection{Photometric Redshift Depth.}
\label{photozdepth}
In previous sections, the \texttt{Odds} parameter was used to retrieve galaxy samples with a common photometric redshift performance. Based on this piece of information, it turns out possible to characterize the so-called photometric redshift depth (hereafter, photo-z depth) of a survey; i.e., to forecast the number of galaxies expected in a survey with a certain photo-z precision, down to a certain magnitude. Although the analysis presented here is described solely in terms of the apparent $r$-band magnitude, the photometric redshift depth of a survey could be described in terms of other variables, such as redshift, morphology, spectral-type, or stellar-mass. 

\vspace{0.2cm}

\begin{figure*}
\includegraphics[width=17.cm]{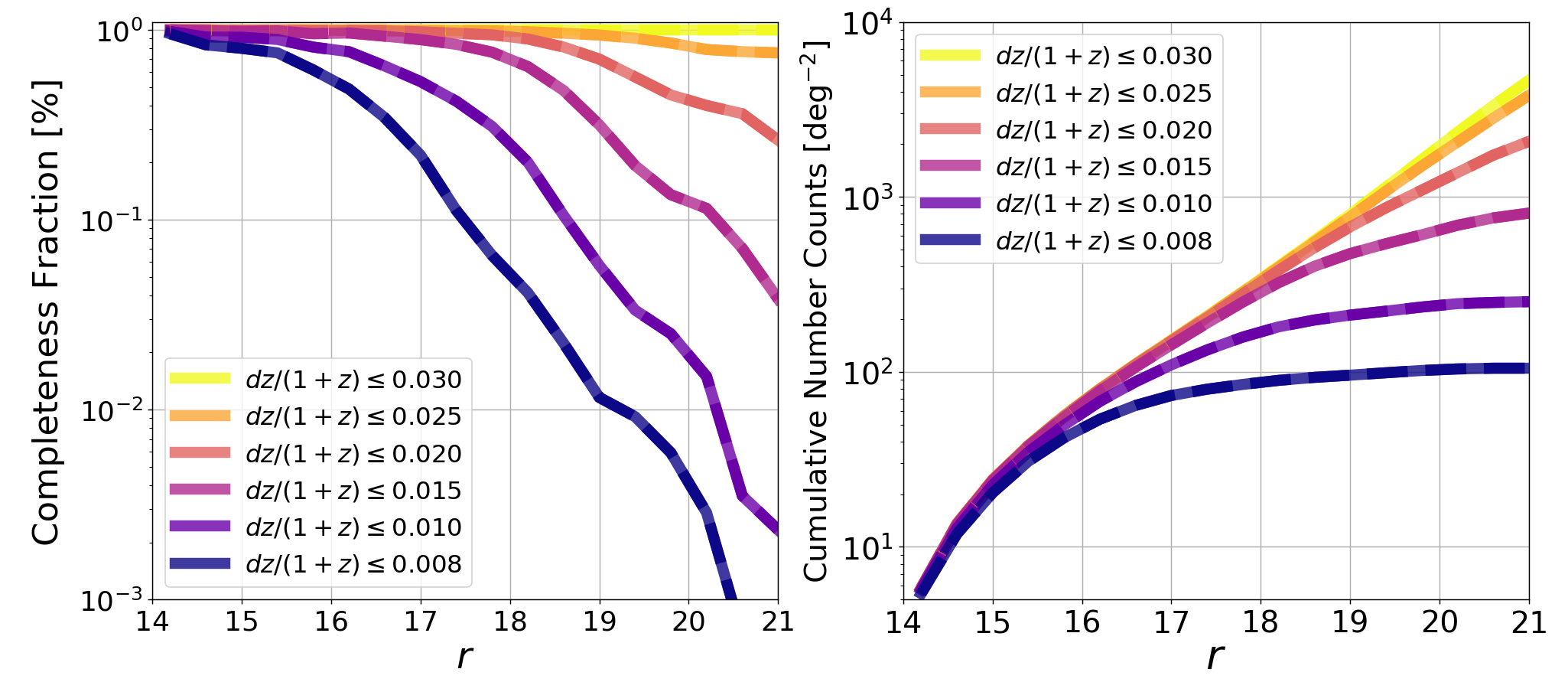}
\caption[]{Photometric Redshift Depth. Left: Per one fraction of galaxies as a function of the $r$-band magnitude with a given photo-z precision. Right: Expected cumulative number of galaxies with a given photo-z precision and magnitude normalized to a 1 degree squared.}
\label{photozcompleteness}
\end{figure*}

In order to estimate the photo-z depth of S-PLUS, we do the following exercise. Initially, we use the information presented in Section \ref{oddsphotoz} and Table \ref{phzacctable0} to define four different \texttt{Odds} cuts (i.e.,\texttt{Odds}$>$0.0, \texttt{Odds}$>$0.4, \texttt{Odds}$>$0.6 \& \texttt{Odds}$>$0.8). This serves to split the photometric sample in groups of galaxies with a common photometric redshift precision: $\delta_{z}/(1+z)<$0.030, $\delta_{z}/(1+z)<$0.020, $\delta_{z}/(1+z)<$0.015 \& $\delta_{z}/(1+z)<$0.010, respectively. Then, we count the number of galaxies per magnitude bin, before and after applying these \texttt{Odds} cuts. Thus, we estimate the fraction of galaxies within different magnitude bins, with a minimum \texttt{Odds} value. The results are illustrated on the left-hand side of Figure \ref{photozcompleteness}, where the completeness fraction of galaxies as a function of the $r$-band magnitude is shown. As seen from the inner panel, the photo-z precision is colour-coded as follows. From bottom to top, different lines correspond to galaxies with a photo-z precision $\delta_{z}/(1+z)<$0.008, $\delta_{z}/(1+z)<$0.010, $<$0.015, $<$0.020, $<$0.025 and $<$0.030, respectively. From this figure we can draw the following conclusions: 80\% of galaxies with a magnitude $r=$20.5 are expected to have a photo-z error $\leq$0.025, 50\% with a photo-z error $\leq$0.020 at a magnitude $r=$19.5 or 50\% with a photo-z error $\leq$0.015 at a magnitude $r=$18.5. Similarly, we find that 1 out 10 galaxies at a magnitude $r=$18.5 is expected to have a photo-z error $\leq$0.01 and that 5 out of 100 galaxies with magnitude $r=$18.0 a photo-z error $\leq$0.008.  

\vspace{0.2cm}

In order to forecast the total expected number of galaxies in S-PLUS with a certain photometric redshift precision, after it completes the observation of the 8000 $deg^{2}$, corresponding the \texttt{MS} region, it is necessary to estimate the expected number of galaxies in S-PLUS per magnitude bin and squared degree. To do so, we have selected all sources classified as galaxies from the 170 fields making the DR1 and computed its average number density as a function of the r-band magnitude; i.e., the averaged number of galaxies per degree squared. Finally, we have combined the previous completeness fraction with the so-estimated expected number of galaxies per magnitude bin in the 8000 $deg^{2}$. The right panel from Figure \ref{photozcompleteness} shows the expected cumulative distribution of galaxies between magnitudes 14$<$r$<$21, where the photo-z precision is colour-coded adopting the previously used criteria. We find that, after S-PLUS completes its observations, a total of $\sim$1M of galaxies with a precision $\delta_{z}/(1+z)\leq$0.008, $\sim$2M galaxies with a $\delta_{z}/(1+z)\leq$0.01, 6.4M with a $\delta_{z}/(1+z)\leq$0.015, 16M with a $\delta_{z}/(1+z)\leq$0.02 and $\sim$32M galaxies with a $\delta_{z}/(1+z)\leq$0.025.    

\vspace{0.2cm}

Finally, these estimates correspond to the expected number of detectable galaxies in our survey which, in practice, does not need to correspond to the total number of galaxies in a given magnitude or redshift bin. As discussed in Section \ref{completABz}, in order to retrieve the real redshift distribution of galaxies in the Local Universe (i.e., $n(z)$), it is necessary first to compute the Completeness matrices as a function of the magnitude and redshift to, afterwards, correct the observed number counts to account for the non-detected galaxies in our images due to selection effects caused by the limited photometric depth of our observations\footnote{Disconsidering other effects like the increase in bright stars at low galactic latitudes and/or regions with high galactic extinction.}.

\subsection{Comparison with other catalogs.}
\label{photozcomparison}

\begin{figure}
\includegraphics[width=8.0cm]{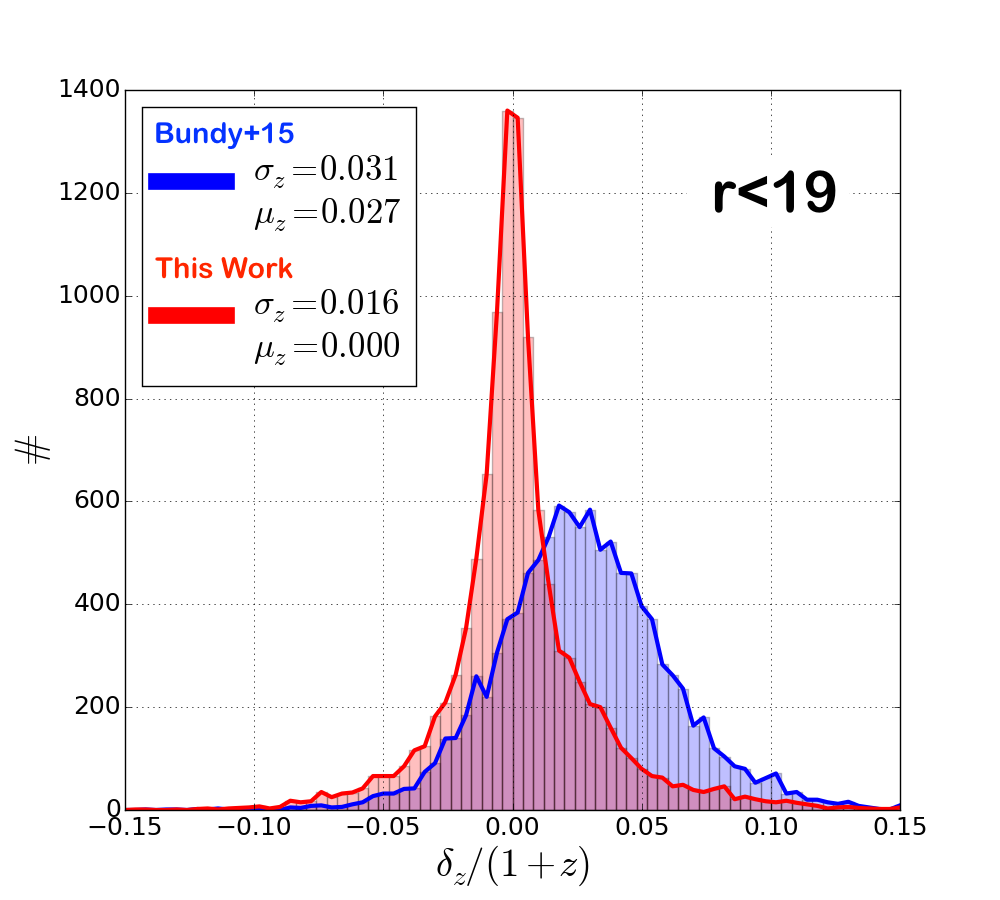}
\caption[photoz Comparison with S82-MGC]{Photometric Redshift Performance compared to S82-MGC.}
\label{photozcompare}
\end{figure}

We take advantage that the Stripe-82 region has been observed by several other astronomical programs which have also performed multi-band photometry and derived photo-z estimates. To demonstrate the benefit of increasing the wavelength resolution (i.e., by including more filters) when estimating photo-z, a sample of $\sim$11k galaxies with magnitudes $r<19.0$ was compiled to compare the performance of the photo-z. Stripe-82 Massive Galaxy Catalog \citep{2015ApJS..221...15B} combines SDSS/ugriz (complete down to a magnitude $r\sim 23.5$AB) and UKIDSS/YJHKs (complete down to a magnitude $r\sim 20$AB) data to derive photo-z in the Stripe-82\footnote{\url{http://www.ucolick.org/~kbundy/massivegalaxies/s82-mgc-catalogs.html}}. Both datasets use the \texttt{BPZ2} code for the redshift estimation. As seen in Figure \ref{photozcompare}, S-PLUS (in red) reaches a precision of $\delta_{z}/(1+z)<$0.016 and a $\mu_{z}=0.000$ for galaxies with a magnitude $r<19.0$. For the exact same sample, \cite{2015ApJS..221...15B} (in blue) reaches a precision of $\delta_{z}/(1+z)<$0.031 and $\mu_{z}=0.027$. Our results corresponds to an improvement of a factor of 2 in accuracy and a factor of 20 in the bias.

\subsection{Importance of the 7 narrow-bands.}
\label{narrowbands}

As demonstrated by many authors (e.g., see Figure B1 from \cite{Molino14}), the reliability of photo-z determinations increases with the number of pass-bands which are used in the computation (see \cite{1994MNRAS.267..911H} or \cite{2009ApJ...692L...5B} for an in-depth discussion). This is specially true in the case of medium-to-narrow pass-bands, since these allow a better sampling of the SED of sources. In this section, we want to quantify the benefit of including 7 additional narrow-bands (hereafter NBs) to classical $u$, $g$, $r$, $i$, $z$ broad-band (hereafter BBs) systems, when computing photo-z estimates as function of the apparent $r$-band magnitude and redshift.

\vspace{0.2cm}

\begin{figure}
\includegraphics[width=8.0cm]{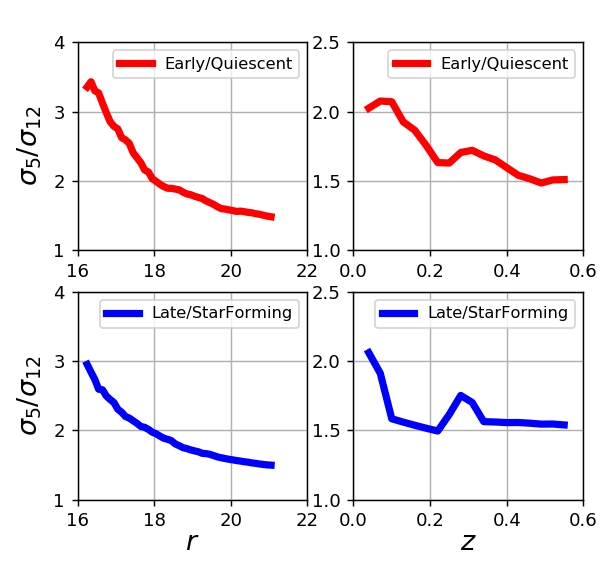}
\caption[5to12bands]{Performance ratio $\sigma_{5}$/$\sigma_{12}$ when computing S-PLUS photo-z using 5 broad-band ($\sigma_{5}$) or 5 broad + 7 narrow-band ($\sigma_{12}$) filters. Ratio as a function of the $r$-band magnitude (left) and redshift (right), for quiescent/early-type (top) and star-forming/late-type galaxies (bottom) galaxies.}
\label{sigmazNBs}
\end{figure}

To cope with this goal, we execute the \texttt{BPZ2} code twice on the dataset presented in Section \ref{photometry} \& Section \ref{controlsample}, adopting the following procedure. Initially, we run \texttt{BPZ2} only with the 5 BB filters, forcing the code to ignore the photometry from the 7 NB filters. Subsequently, we rerun \texttt{BPZ2} again but letting the code to use the entire dataset. Since the setting of \texttt{BPZ2} is kept the same in both runs, the differences in the final performance simply reflect the importance of the wavelength-resolution when mapping out the SED of sources. The neat improvement in the redshift estimation given by the NBs is presented in Figure \ref{sigmazNBs}. In order to facilitate this comparison, we prefered to compute the ratio among precision $\sigma_{5}$/$\sigma_{12}$.  

\vspace{0.2cm}

On the one hand, as shown in the left panel of Figure \ref{sigmazNBs} where the precision is estimated separately for early- and late-type galaxies as a function of the $r$-band magnitude, including the additional 7 NBs leads to an improvement of a factor of 4 for galaxies with magnitudes $r<15$, a factor of 2.5 for magnitudes $15<r<17$ and/or a factor of 1.7 for magnitudes $17<r<19$. This reflects the fact that when the signal-to-noise of the detections is high, photo-z estimates can dramatically improve those from classical systems. On the other hand, as shown in the right panel where the precision is estimated as a function of the redshift ($z$), a factor of 2 improvement is found for galaxies with $z<0.1$ and a factor of 1.5 for $0.1<z<0.4$. Again, these results illustrate the enormous benefit of including additional narrow-bands to standard photometric systems. Interestingly, as pointed out in \cite{Molino18}, surveys such as SDSS, KiDS or DES based on standard broad-bands, cannot surpass a certain precision in their photo-z estimates irrespectively of the signal-to-noise of their observations. This limiting-factor comes from the limited wavelength resolution provided by the broad-bands, which causes a degeneracy in the colour-redshift space.  

\subsection{Redshift window opportunities.}
\label{zWindows}

As discussed in the Section \ref{narrowbands}, photometric redshift estimates computed from standard $ugriz$ broad-band filter systems can be largely improved if they are complemented by medium-or-narrow pass-bands. Whereas broad-band filters mainly serve to constrain the continuum of the SED of sources, narrow-band filters allow the detection of other spectral-features (such as emission or absorption lines), which help to break (or to reduce) the colour-redshift degeneracies and, therefore, to downsize the photometric redshift uncertainties. As discussed in \cite{jpasredbook}, the new generation of photometric redshift surveys (such as J-PAS \citep{jpasredbook} and PAU \citep{2014MNRAS.442...92M}, will utilize optimized filter systems made as a combination of broad and narrow pass-bands to get the best of each world and so maximize its performance at all magnitude and redshift ranges.

\vspace{0.2cm}

\begin{figure}
\includegraphics[width=8.7cm]{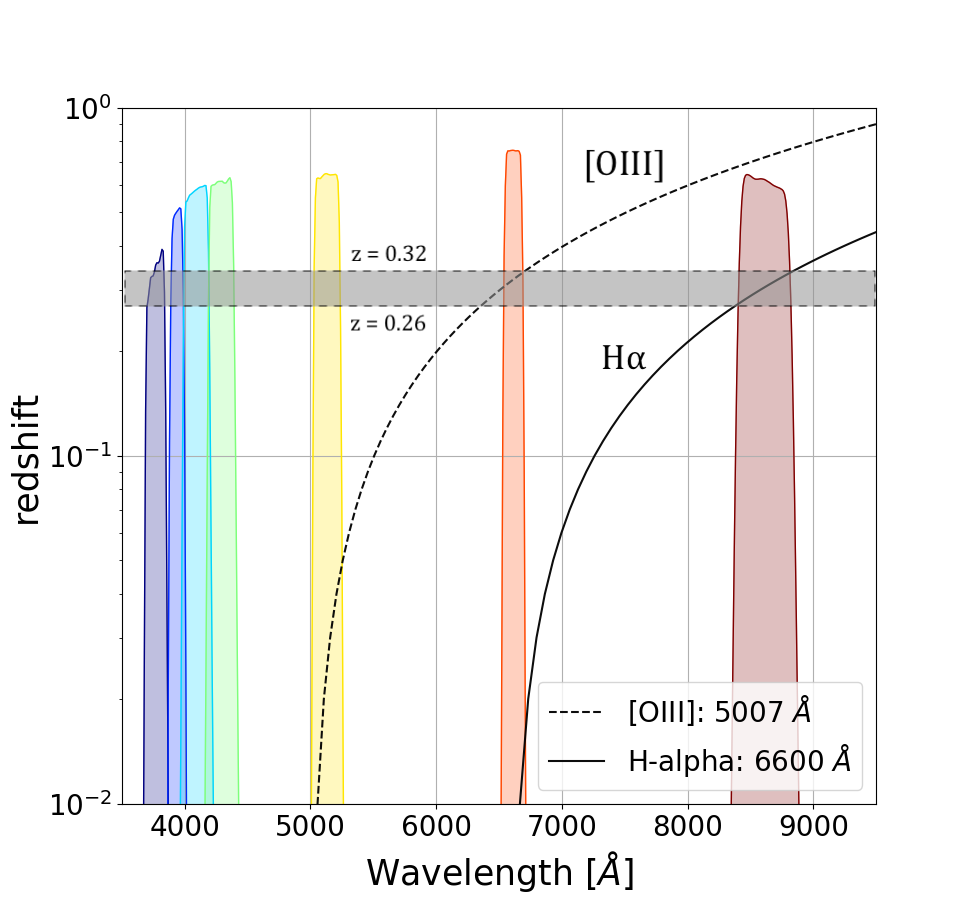}
\includegraphics[width=8.2cm]{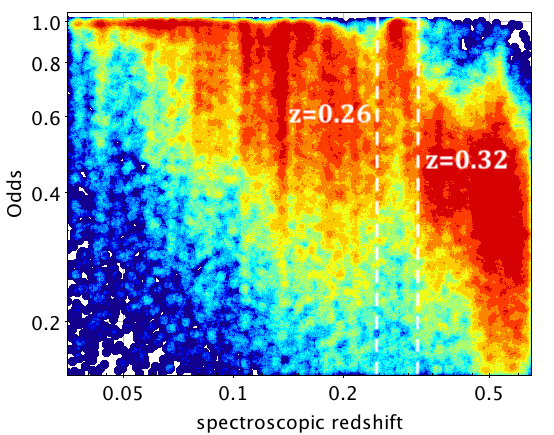}
\caption[]{Redshift window opportunity. Top: The figure shows the simultaneous detection of the [OIII] and H$\alpha$ emission-lines, by the $J0660$ \& $J0861$ narrow-band filters at a redshift interval 0.26$<z<$0.32. Botton: Distribution of the \texttt{Odds} values as a function of the redshift, for the $\sim$100k galaxies in the spectroscopic control sample. It is observed an upturn in the distribution at the aforementioned redshift interval. This very effect represents a boost in the photo-z estimates, bringing an opportunity to conduct statistical analysis.}
\label{zwindows}
\end{figure}

It is worth mentioning that the number and wavelength distribution of these narrow-band filters in a filter system will define a set of redshift windows within which a survey might be able to detect specific spectral features from astronomical sources\footnote{If the signal-to-noise is large enough to allow the detection at that magnitude or redshift range.}. In this section, we cope with this goal, finding the redshift windows defined by the S-PLUS filter system, by relying on the \texttt{Odds} parameter since it encodes the performance of our estimates in a rather simple manner. As illustrated in the top panel of Figure \ref{zwindows}, where the wavelength evolution of the emission-lines [OIII] and H$\alpha$ is shown as a function of redshift, there exist a redshift interval (0.26$<z<$0.32) where these emission-lines enter simultaneously both the $J0660$ and $J0861$ narrow-band filters, respectively. 

\vspace{0.2cm}

This fact causes the photometric redshift estimates to be more reliable and, therefore, to increase its \texttt{Odds} value. In the bottom panel of Figure \ref{zwindows}, we represent the distribution of the obtained \texttt{Odds} values as a function of the redshift for all galaxies in our spectroscopic control sample. Colours represent the number density, being red densely and blue sparsely populated areas. As expected, the \texttt{Odds} distribution gets values close or equal to 1 for galaxies at $z<0.15$ and declines steadily to lower values at $z>0.15$. This behaviour is expected since most galaxies at low-z are detected with high signal-to-noise and high redshift galaxies tend to have a noisier photometry. Interestingly, in the exact redshift interval where the two emission-lines mentioned before are supposed to be simultaneously detected by the narrow-band filters (0.26$<z<$0.32), the \texttt{Odds} shows a clear upturn, passing from the expected \texttt{Odds}$\sim$0.6 to \texttt{Odds}$\sim$1.0. This increase in the \texttt{Odds} value can be translated, according to the discussion stated in Section \ref{oddsphotoz}, into an improvement in the photo-z precision. Complementary to this discussion, as demonstrated in Figure \ref{sigmazNBs} of Section \ref{narrowbands}, this effect is only observed when we include the 7 narrow-band filters in our filter system. This means that this redshift window is a distinctive feature of the S-PLUS survey, and might represent an opportunity to conduct statistical analysis where photo-z precision and cosmological volume is required, such as luminosity or mass functions. 

\section{Probability Density Functions}
\label{PDFs}

As stressed in \cite{2015MNRAS.452.3710R}, in order to enter the era of precision cosmology \citep{2006astro.ph..9591A}, one must be able to incorporate the uncertainty in the redshift estimate into any cosmological analysis. This statement highlights the importance of stop treating photometric redshifts as simple point-estimates and start thinking of them as multi-dimensional PDFs.

\vspace{0.2cm}

As emphasized in Section \ref{bpzcode}, when we introduced the $\texttt{BPZ2}$ code, in this work we have computed the full PDF in a bi-dimensional redshift-spectral-type space, for every source detected in our images. Similar to what we did in Section \ref{testing} to evaluate the performance of our photo-z estimates as if they were simple point-estimates, in this section we instead make use of the entire PDFs in both redshift and spectral-type space. Nowadays, there are an increasing number of works where it is emphasized not only the benefit of treating photometric redshift estimates as PDFs rather than as simple point-estimates, but also giving recipies about how to characterize the reliability of these distributions encoding photo-z uncertainties along with different approaches to compensate underestimated (or overestimated) PDFs (\citealt{2000ApJ...536..571B}; \citealt{2002MNRAS.330..889F}; \citealt{2006AJ....132..926C}; \citealt{2008MNRAS.386..781M}; \citealt{2009MNRAS.396.2379C}; \citealt{2009A&A...508.1173P}; \citealt{2009ApJ...700L.174W}; \citealt{2010MNRAS.406..881B}; \citealt{2011ApJ...734...36A}; \citealt{2012ApJS..201...32S}; \citealt{2013MNRAS.432.1483C}; \citealt{Molino14}; \citealt{2014MNRAS.438.3409C}; \citealt{2014MNRAS.442.3380C}; \citealt{2015A&A...576A..53L}; \citealt{2015A&A...576A..25V}; \citealt{2017A&A...599A..62L}; \citealt{Molino18}; \citealt{2018MNRAS.475..331G}, among others) 

\vspace{0.2cm}

The analytical tools needed to characterize the performance of these distribution functions differ from those previously utilized in Section \ref{testing}. Through the following section, we introduce and utilize a number of approaches to quantify the reliability of our PDFs estimations using the \texttt{BPZ2}, encoding the real photo-z uncertainties. 

\subsection{Measuring Confidence Levels}
\label{HPD}

We measure the reliability of our PDFs encoding real uncertainties in photometric redshift estimates using the Highest Probability Density (HPD) technique, as described in \cite{2016MNRAS.457.4005W}. This statistical method is based on the Quantile-Quantile (Q-Q) plots, where the distribution of threshold credible intervals, \texttt{C}, is calculated from a spectroscopic redshift sample. This approach assumes that if PDFs properly represent the redshift uncertainty, the expected distribution of \texttt{C} values should be constant between 0 and 1, with the cumulative distribution function \texttt{$\hat F(C)$} (or CDF) following a 1:1 relation as in a quantile-quantile plot (Q-Q). This technique, which has been implemented in a number of works (i.e., \cite{2017MNRAS.468.4556F}, \cite{2017ApJ...838....5L}, \cite{2017MNRAS.465.1959C}, \cite{2018MNRAS.475..331G} or \cite{2018MNRAS.473.2655D}, among others), has proven to be very efficient in describing over- or under-confidence. For example, stressing whether a PDF departs from Gaussianity due to the presence of heavier tails or a larger skew. In addition, we also investigate an optimization Kernel to calibrate our PDFs. This analysis is divided in four categories: magnitude, redshift, spectral-type and \texttt{Odds}.

\vspace{0.2cm}
 
To understand the reliability of our PDFs for all types of galaxies in our catalogs, we divide the spectroscopic redshift sample in multiple intervals: in magnitude bins ($r<17$, $r<18$, $r<19$, $r<20$ and $r<21$), in redshift bins ($z<0.1$, $0.1<z<0.2$, $0.2<z<0.3$, $0.3<z<0.4$ and $0.4<z<0.5$), in \texttt{Odds} bins (\texttt{Odds}$>$0.0, \texttt{Odds}$>$0.3, \texttt{Odds}$>$0.6 and \texttt{Odds}$>$0.9) and in spectral-types (separating our SED models in Early- and Late-types). We explore the dependence of the \texttt{$\hat F(C)$} function as a function of each of the before-mentioned bins. According to the HDP test, we find that our PDFs are systematically underestimated, with a deviation larger the brighter the galaxy, the lower the redshift and the lower the \texttt{Odds} value. Likewise, we find that late-type galaxies show a larger deviation than early-types. 

\vspace{0.2cm}

\begin{figure}
\includegraphics[width=8.2cm]{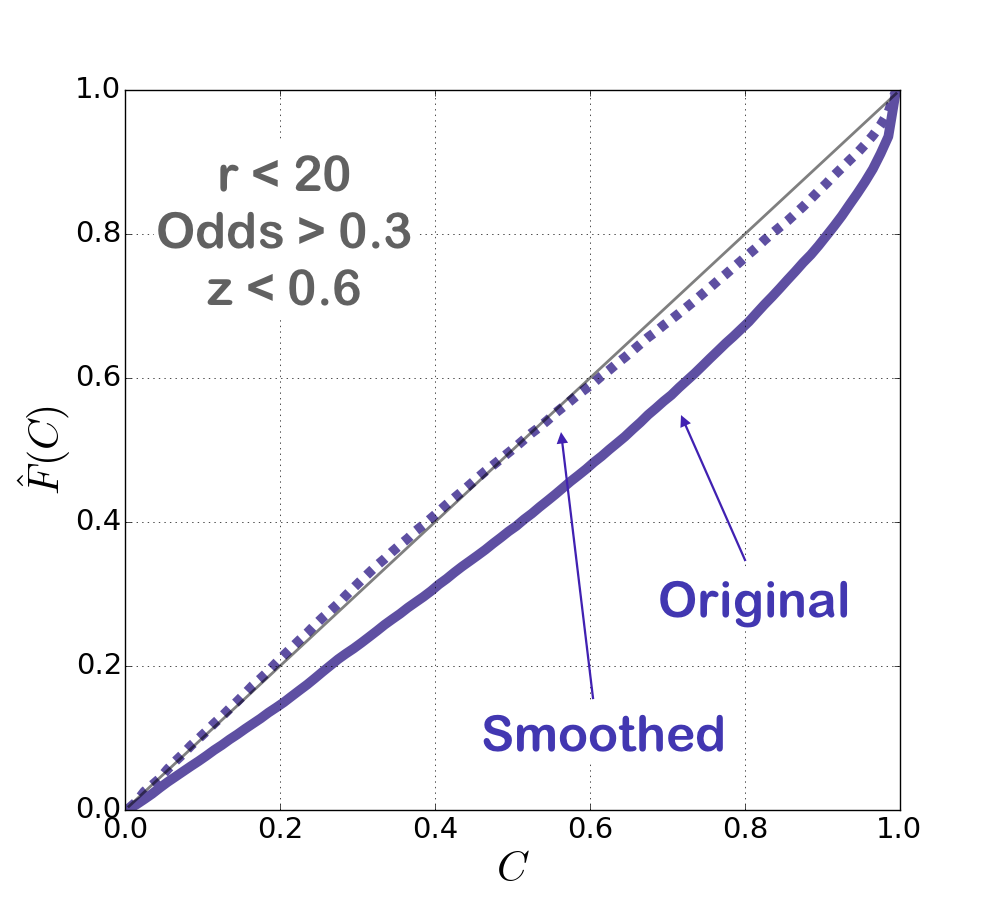}
\caption[The quantile-quantile (i.e., Q-Q) \texttt{$\hat F(C)$} plot.]{We measure the reliability of our PDFs encoding real uncertainties using the Highest Probability Density (HPD) technique. We recommend to apply a smoothing Gaussian-Kernel of equivalent width $\sigma_{GK}$=0.019 to our PDFs, in order to avoid possible under-estimation effects.}
\label{Wittman15Test}
\end{figure}

In order to compensate this bias, we look for an optimal Gaussian Kernel ($\sigma_{GK}$) to be convolved with our raw PDFs to bring them to the desired 1-to-1 line. In this exercise we explore a range of values from 0.005$<\sigma_{GK}<$0.03. Although it would be ideal to apply an optimal GK to each individual galaxy according to its magnitude, redshift, \texttt{Odds} value and most likely spectral-type, this approach is computationally expensive. Instead, we prefer to adopt a simpler approach, defining a unique Gaussian Kernel which represents a good compromise between accuracy and simplicity. After a careful examination of the aforementioned parameter space, we recommend to apply a $\sigma_{GK}$=0.018 to our PDFs. As shown in Figure \ref{Wittman15Test}, this GK will assure that most galaxies with good photometric redshift estimates in our catalogs (i.e., those with a magnitude $r<$20, a redshift $z<$0.6 and \texttt{Odds}$>$0.3) will have reliable PDFs for statistical analysis.  

\vspace{0.2cm}

\subsection{Encoding photometric uncertainties (II).}
\label{PDFdz}

\begin{figure*}
\includegraphics[width=16.0cm]{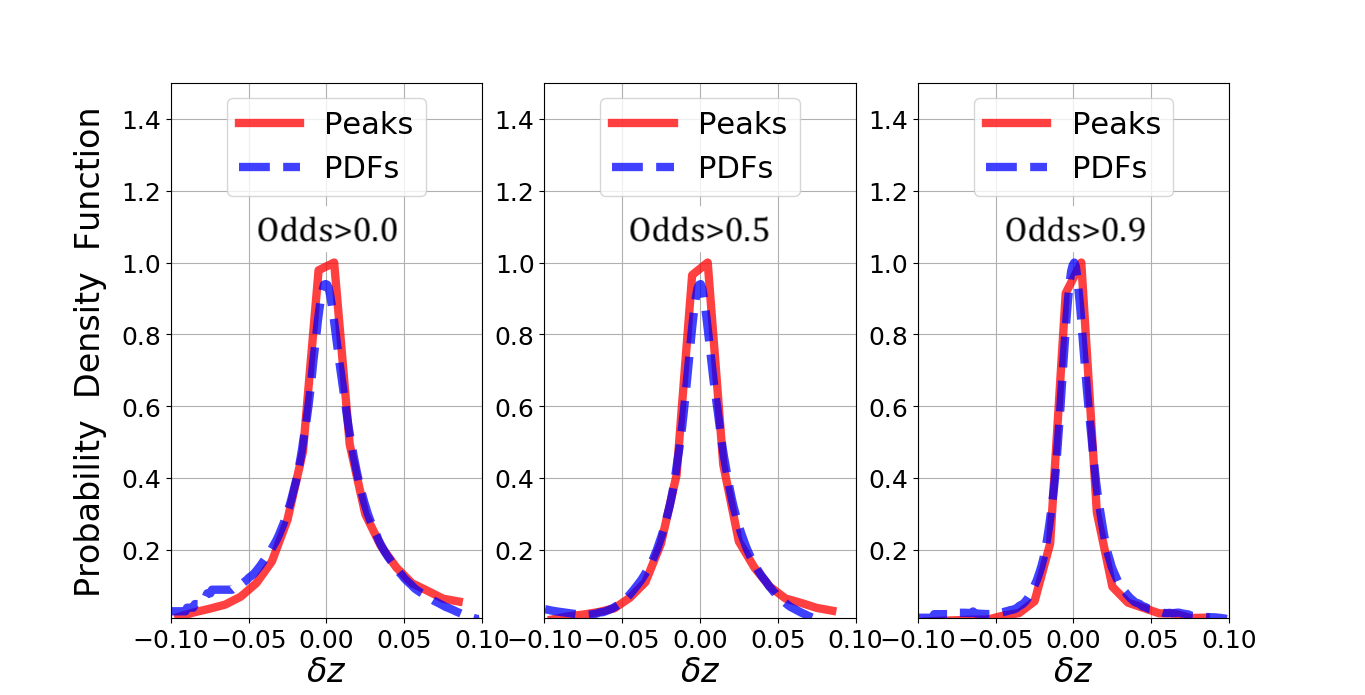}
\caption[Reliability of PDF]{In this figure we show the reliability of the PDFs encoding the photometric redshift uncertainties. From left to right, the distribution correspond to galaxies with magnitude r$<$21 and \texttt{Odds}$>$0.0, \texttt{Odds}$>$0.5 \& \texttt{Odds}$>$0.9.}
\label{dzPDF}
\end{figure*}

In this Section, we present a complementary discussion about the reliability of the PDFs encoding  photometric-z uncertainties. We compare the error distribution functions obtained from our photometric redshift estimates (see Section \ref{metric}), when they are treated as single-point estimates or as probability distribution functions (i.e., PDFs). In the latter case, before the stacking it is necessary to normalize individual PDFs and subtract the spectroscopic redshift value from each galaxy. We do not separate here galaxies by their spectral-types. As in previous analysis, we divide our spectroscopic redshift galaxy sample in three different \texttt{Odds} bins. We also select galaxies with magnitudes 16$<$r$<$21 to assure a well-behaved photometry avoiding very bright and very faint sources. As seen in Figure \ref{dzPDF}, where the error distribution from point estimates is represented in red and that from the stacked PDFs in blue, both distributions are in good agreement. As expected, galaxies with low \texttt{Odds} values (e.g., left panel) tend to show larger tails than those from point-like estimates. The excess signal comes from secondary peaks in the distribution functions; information ignored by single-point estimates. Galaxies with higher \texttt{Odds} values (e.g., intermediate and right panel), show narrow distribution with little excess in the winds. This effect reflects the fact PDFs for galaxies with high \texttt{Odds} are mostly described by single-peak distributions.   

\subsection{Reliability mapping the n(z)}
\label{nzspec}

As explained in Section \ref{PDFs}, the $\texttt{BPZ2}$ code computes a bi-dimensional redshift vs spectral-type probability distribution function for every source in our catalogues. By means of a simple marginalization over the spectral-type information, these bi-dimensional distributions can be collapsed into a single 1-dimensional distribution (i.e., redshift) space that we named here as zPDF. These distribution functions need to satisfy the following normalization criteria to preserve its dimensionality:

\begin{equation}
zPDF_{i} = {\int_{T}} p_{i}(z,T|D) \,dT = {\int_{z}} p_{i}(z|D) \,dz = 1  
\label{zPDFnorm1}
\end{equation} 

\vspace{0.1cm}

where $p_{i}(z,T|D)$ represents the probability distribution function in both redshift ($z$) and spectral-type ($T$) space for the $i$th-galaxy and $p_{i}(z|D)$ the collapsed probability distribution function in redshift after marginalize over templates. 

\vspace{0.2cm}

\begin{figure*}
\includegraphics[width=16.0cm]{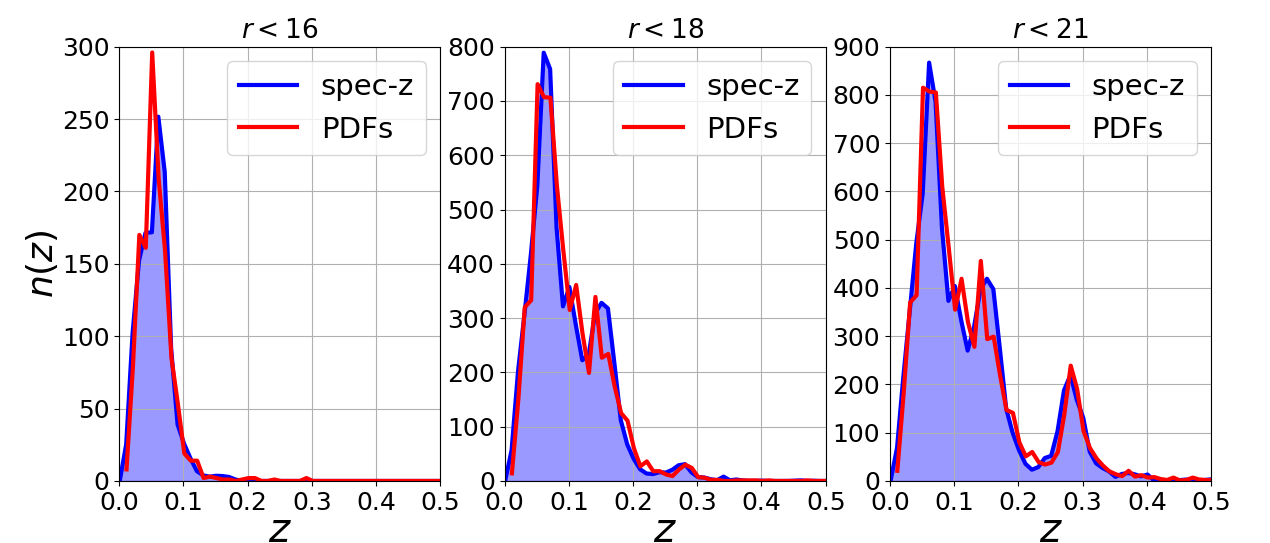}
\caption[]{Comparison between the spectroscopic redshift distribution (blue) and the stacked photo-z PDF (red), for a sample of galaxies with $\texttt{Odds} \geq0.9$ and magnitudes $r<$16, $r<$18, $r<$21. A redshift resolution $\Delta$z=0.01 is adopted to facilitate its visualization.}
\label{ndzPDF1}
\end{figure*}

In this section, we briefly illustrate the capability of our PDFs retrieving the redshift distribution of galaxies (i.e., $n(z)$) in the nearby Universe. We refer the interested reader to \textcolor{blue}{L\'opez-Sanjuan et al., (in prep.)} for an in-depth discussion on the subject. To do so, we compute the PDF-based redshift distribution for all the galaxies with a spectroscopic redshift value lower than $z<0.5$ and a $\texttt{Odds}$ value $\geq0.9$, to select galaxies with very secure photometric redshifts in the nearby Universe. The so-selected sample is further divided in three magnitude bins (i.e., $r<$16, $r<$18, $r<$21), to figure the performance of these distributions with the magnitude of sources. For the sake of simplicity, we have preferred not to split the  sample according to their spectral-type classification (e.g., among red/blue galaxies) but treating it as a single population. Finally, a redshift resolution of $\Delta$z=0.01 has been adopted to facilitate its visualization. In Figure \ref{ndzPDF1} we compare the PDF-based photometric redshift distribution (red) with the spectroscopic redshift one (blue). As seen from this figure, our $zPDF$ can successfully retrieve the real distribution of galaxies in the nearby Universe; opening the possibility of revisiting the redshift distribution of galaxies in the nearby Universe down to a magnitudes $r<$21.  

\subsection{PDFs for stellar-sources}
\label{starPDFs}
The S-PLUS photometric catalog includes a Machine Learning-based statistical star/galaxy classification, computed using colours and apparent morphology from sources (\textcolor{blue}{Costa-Duarte et al., (in prep.)}). Due to the statistical nature of this methodology, every single source in our catalog has been associated with a probability of being a star or a galaxy (see \textcolor{blue}{Sampedro et al., (in prep.)} for more details). Therefore, the \texttt{BPZ2} code is run on every detected source independently of its true nature. In this section, we investigate how stars misclassified as being galaxies may contaminate extragalactic analysis. In particular, we tackle this issue by making use of the redshift Probability Distribution Function (i.e., zPDF) of sources, since these distributions are further recommended for large statistical analysis since they encode more information from sources that classical point-estimates (see \cite{2017A&A...599A..62L} for a longer discussion). 

\vspace{0.2cm}

\begin{figure}
\includegraphics[width=8.cm]{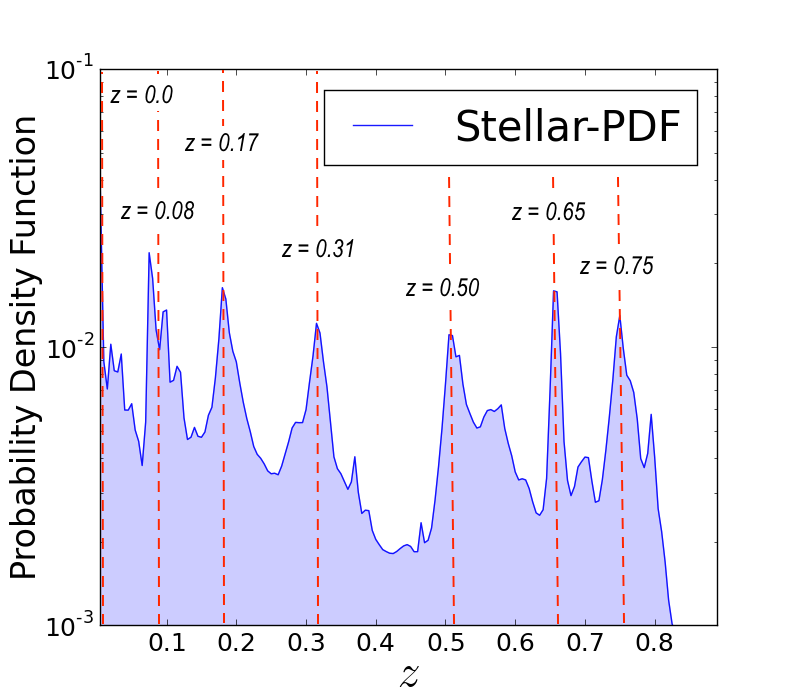}
\caption[Redshift Distribution of Stellar-sources]{Redshift Probability Density Function (zPDF) for stellar sources. The figure shows the regions where misclassified galactic sources as galaxies may contaminate the most photometric redshift distributions.}
\label{PDFstars}
\end{figure}

In order to understand how the zPDF of misclassified stars may look like, we did the following exercise. We cross-matched the SDSS/S82 stellar catalog (\cite{2007AJ....134..973I}) with our observations, finding $\sim$250k stars down to a magnitude $r$-band = 21. On these sources, we run the \texttt{BPZ2} code using the S-PLUS multi-band photometry and derive the corresponding zPDF. Finally, we stacked and normalized the final distribution which we named as ``Stellar-PDF". As illustrated in Figure \ref{PDFstars}, there exists several redshift windows at which photometric redshift estimates for stellar sources tend to cluster. As expected, the most prominent peak sits at redshift $z$=0.0 but there are other secondary peaks at $z$=0.008, $z$=0.17, $z$=0.31, $z$=0.50, $z$=0.61 and $z$=0.75, among others. Therefore, these regions are more sensitive to include misclassified stars (as galaxies), contaminating any extragalactic analysis. Interestingly, the redshift window described in Section \ref{zWindows} (i.e., 0.26$<z<$0.32), show a minimum in the Stellar-PDF, reassuring the usability of this redshift interval for scientific studies.   

\section{Absolute Magnitudes and Stellar Masses.}
\label{absmagStellMass}
Through this section, we describe how Absolute Magnitudes (\S\ref{Mabs}) and Stellar Masses (\S\ref{STM}) have been estimated for our galaxies, and how photometric redshift uncertainties may affect the estimation of those quantities. 

\subsection{Absolute Magnitudes.}
\label{Mabs}

\begin{figure}
\includegraphics[width=7.5cm]{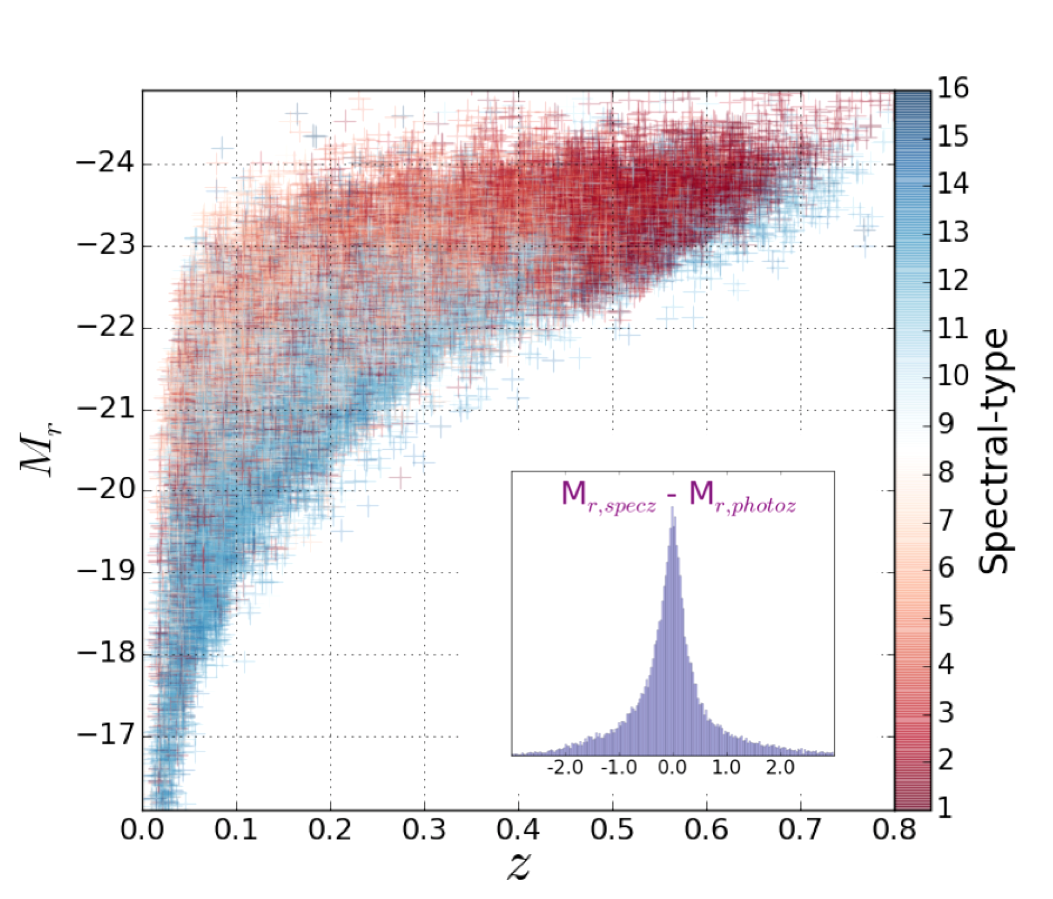} 
\caption[]{Absolute Magnitude ($M_{r}$) versus redshift for galaxies detected in the Stripe-82, according to \texttt{BPZ2} code. The most likely spectral-type of galaxies is colour coded as indicated by the vertical colour-bar. Inner panel shows the logarithmic error distribution observed when computing the Absolute Magnitudes with and without fixing the redshift of galaxies, which has a typical 1-$\sigma$ dispersion of RMS = 0.5 magnitudes.}
\label{AbsMrSpecz}
\end{figure}

We use the \texttt{BPZ2} code to compute the Absolute Magnitudes in the $r$-band for all sources, according to the most likely redshift and spectral-type, based on the S-PLUS multi-band photometry. These estimates include a template-dependent K-correction described in Table \ref{bpzkcorrections}. In order to understand the quality of our estimates, we study the impact of the photometric redshift uncertainties when deriving Absolute Magnitudes. To do so, initially we force \texttt{BPZ2} to use the spectroscopic redshift value for each galaxy previous to the computation of the absolute magnitude. Redshifting the SED models to the exact redshift of each individual galaxy, letting \texttt{BPZ2} just to look for the model that best fits the data, renders possible to minimize the uncertainties over the spectral-type classification; i.e., breaking down the colour-redshift degeneracy. After \texttt{BPZ2} finds the best template, it computes the corresponding Absolute Magnitude. As stated before, the comparison of both distributions (i.e., the one using spectroscopic and the one using photometric redshifts) serves to quantify the impact of redshift uncertainties when computing absolute magnitudes for all galaxies in the S-PLUS catalogs. In Figure \ref{AbsMrSpecz}, we show the redshift versus absolute magnitude in the $r-$band (i.e., $M_{r}$) for the sample of spectroscopic galaxies presented in Section \ref{controlsample}. As indicated by the vertical label, the spectral-type of galaxies has been colour-coded; where early-type galaxies appear with red colours and late-type galaxies with blue ones. As expected, the most luminous galaxies in our spectroscopic sample correspond to Luminous Red Galaxies (i.e., LRGs), which start dominating the sample at a redshift $z>0.4$. Inner panel shows the error distribution observed when computing the Absolute Magnitudes with and without fixing the redshift of galaxies, which has a typical 1-$\sigma$ dispersion of RMS = 0.5 magnitudes. Table \ref{bpzkcorrections} includes a complete description of \textit{k}-corrections which have been performed by the \texttt{BPZ2} code.

\subsection{Stellar-Mass.}
\label{STM}

We rely on the \texttt{BPZ2} code to compute stellar masses which uses a mass-to-light empirical ration derived from a colour-magnitude relation similar to that presented in \cite{2011MNRAS.418.1587T} for the GAMA survey\footnote{\url{http://www.gama-survey.org}}, however refined by \cite{2018arXiv180503609L} based on the ALHAMBRA survey\footnote{\url{http://www.alhambrasurvey.com}}\footnote{\url{http://cosmo.iaa.es/content/ALHAMBRA-Gold-catalog}} data. According to the authors, the expected uncertainty in these estimates is of the order of $\sigma$=0.02dex for red galaxies and $\sigma$=0.06dex for late-type galaxies. For the sake of simplicity, here we include the mass-to-light relations (i.e., presented in the aforementioned paper) for red/early-type (Eq. \ref{stmassET1}) and for blue/late-type galaxies (Eq. \ref{stmassET2}):

\begin{equation}
log(M_{\odot}/L_{i})_{r} = 1.02 + 0.84 \times (g-i) 
\label{stmassET1}
\end{equation}

\begin{equation}
log(M_{\odot}/L_{i})_{b} = 1.41 + 0.21 \times (g-i) + 0.14 \times (g-i)^{2} 
\label{stmassET2}
\end{equation}

\vspace{0.2cm}

\begin{figure}
\includegraphics[width=7.5cm]{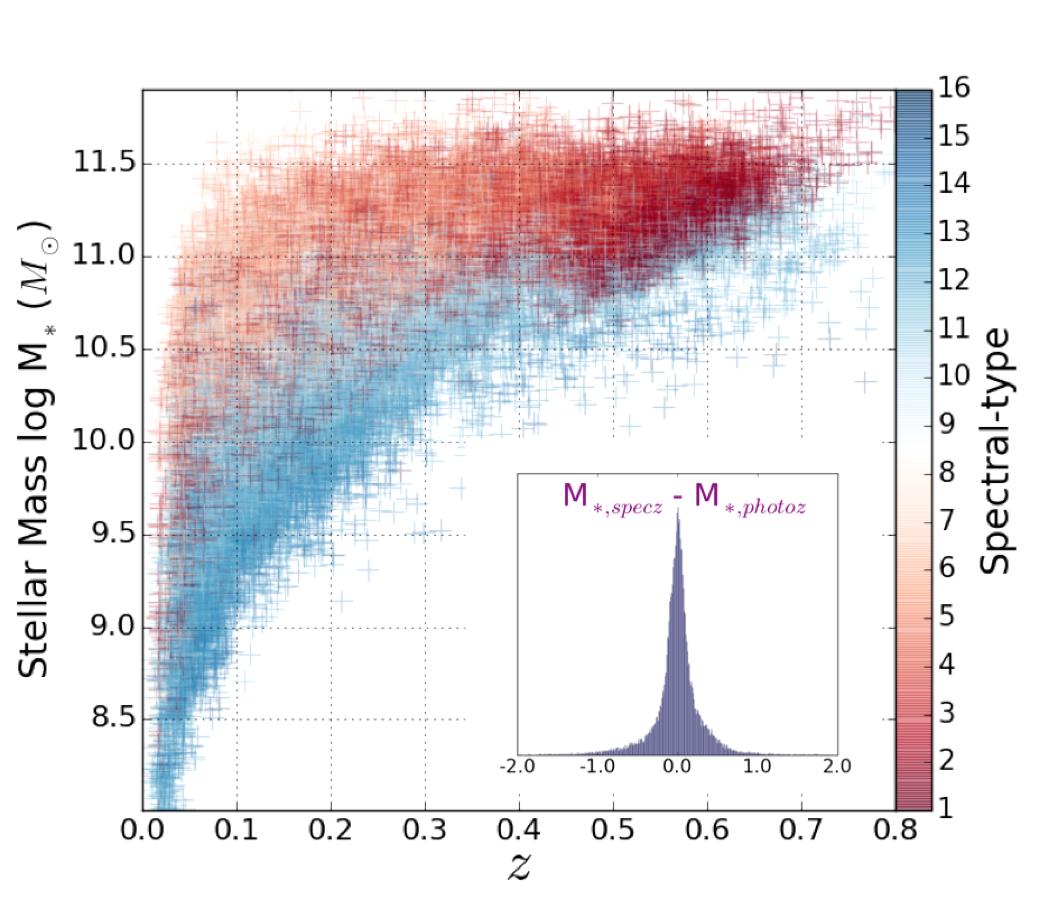}
\caption[]{Stellar-Mass ($M_{*}$) versus redshift for galaxies detected in the Stripe-82, according to \texttt{BPZ2} code. The most likely spectral-type of galaxies is colour coded as indicated by the vertical colour-bar. Inner panel shows the logarithmic error distribution observed when computing the stellar-mass with and without fixing the redshift of galaxies, which has a typical 1-$\sigma$ dispersion of RMS = \textbf{0.1} magnitudes.}
\label{sm1}
\end{figure}

As in the previous section, we study the impact of the photometric redshift uncertainties when computing stellar-masses and compare our results with those from the literature. In Figure \ref{sm1}, we show the redshift versus stellar-mass (i.e., $M_{*}$) for the sample of spectroscopic galaxies presented in Section \ref{controlsample}. As indicated by the vertical label, the spectral-type of galaxies has been colour-coded; where early-type galaxies appear with red colours and late-type galaxies with blue ones. As expected, the most luminous galaxies in our spectroscopic sample correspond to Luminous Red Galaxies (i.e., LRGs), which start dominating the sample at a redshift $z>0.4$. Inner panel shows the logarithmic error distribution observed when computing the stellar-mass with and without fixing the redshift of galaxies, which has a typical 1-$\sigma$ dispersion of RMS = \textbf{0.1} magnitudes. 

\section{Testing the S-PLUS photometry.}
\label{testingphoto}

In this section, we present a number of tests aiming at validating the quality of our photometric catalogs. Initially, in Section \ref{zpoff}, we use the spectroscopic redshift sample to check the quality of the photometric zero-point estimates. Finally, in Section \ref{noise}, we take advantage of using a SED-fitting based photo-z code to characterize the accuracy of the photometric uncertainties provided for sources in our catalogs.  

\subsection{Photometric ZP-offsets}
\label{zpoff}

As explained in Section \ref{bpzcode}, one of the advantages of running SED-fitting based photometric redshift codes on a sample of galaxies with spectroscopic redshift information is the possibility of making comparisons between the observed and the predicted colours of galaxies. These comparisons serve to a double purpose. When these quantities are computed from a large sample of heterogeneous galaxies at different magnitudes and redshifts, it becomes possible to flag systematic zero-point offsets coming from the initial photometric calibrations. As demonstrated in many works (e.g., \citealt{2006AJ....132..926C}, \citealt{2014A&A...562A..86J}, \citealt{Molino14}) these corrections may enhance the overall photometric redshift precision since they improve the agreement between data and models. Likewise, when these differences are represented as a function of the magnitude (or the signal-to-noise) for each individual band, these error distributions may warn about systematics related to the PSF-homogenization across filters or issues with the electronic response of the CCD camera.

\vspace{0.2cm}

\begin{figure}
\includegraphics[width=7.5cm]{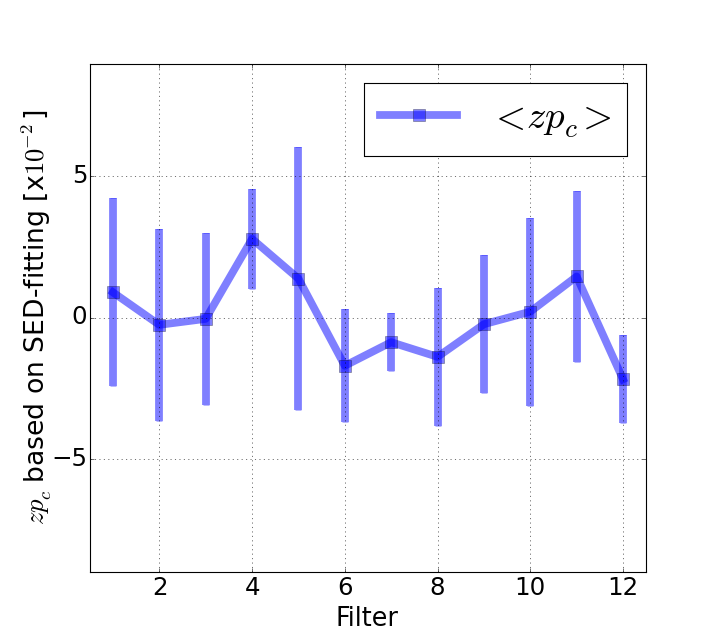}
\caption[Photometric Zero-Point Corrections based on SED-fitting.]{The figure shows the computed ZP-corrections to our photometry according to \texttt{BPZ2}, based on the spectroscopic redshift sample.}
\label{zpcorSEDs}
\end{figure}

Based on the aforementioned ideas, we utilized the spectroscopic redshift sample presented in Section \ref{controlsample}, to look for potential systematics in the S-PLUS multi-band photometry (\textcolor{blue}{Sampedro et al., (in prep.)}). Since the S-PLUS/Stripe-82 observations are made of 170 different pointings, the calibration process is first run individually in each field and then combined into a final averaged value. Initially, we start by comparing the so-derived zero-point corrections as a function of wavelength (i.e., filters). The results from this exercise are illustrated in Figure \ref{zpcorSEDs}. We find small deviations in every filter, smaller than a few hundredth of a magnitude. Interestingly, except for the fourth and twelfth filter, the intrinsic dispersion is always larger than the corrections. This fact makes it complicated to assure these offsets correspond to real issues in the photometry more than to the intrinsic photometric dispersion of galaxy colours due to the noise. The observed zero-point corrections are noted down in Table \ref{tablezps}.

\vspace{0.2cm}

\begin{figure}
\includegraphics[width=7.5cm]{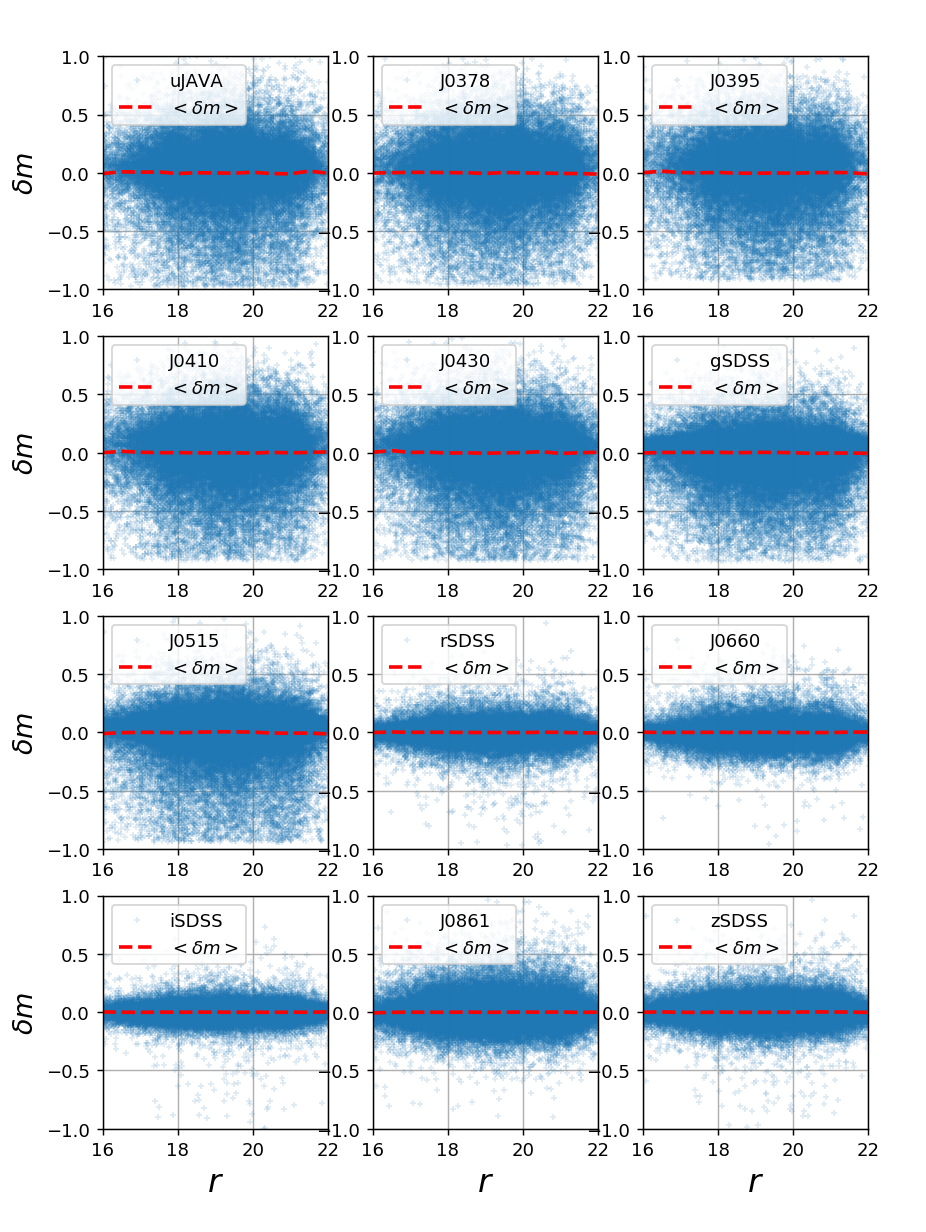}
\caption[Magnitude-dependent Photometric ZP Refinement.]{Comparison between the observed and the expected magnitudes as a function of the r-band magnitude. Internal red lines correspond to the average value binned in magnitude bins of $\delta$m = 0.1. These distributions correspond to the global values for each filter averaged over all independent fields making up the S-PLUS/Stripe-82 catalog. No systematic is observed neither for bright nor for faint sources, assuring the reliability of the S-PLUS photometry.}
\label{magsZPs}
\end{figure}

Later on, we represent these zero-point corrections as a function of the magnitude. In this case, we look for systematics in the way the photometry is performed. As before, we combine the results from the 170 individual fields. Figure \ref{magsZPs} shows an example of the observed differences between predicted and observed colours as a function of the r-band magnitude. Internal red line corresponds to the average value binned in bins of $\delta$m = 0.1 magnitudes. This result proves that there seem not to be any dependence with the magnitude (or surface brightness), assuring that the photometry is robust even for faint sources. 

\begin{table}
\caption{{\small \textbf{S-PLUS Photometric re-calibration.} The table summarizes the main photometric zero-point refinements derived with the \texttt{BPZ2} using galaxies with spectroscopic redshift. It includes the filter name ($Filter$), the average zero-point corrections ($<ZP_{off}>$) and average zero-point dispersion ($\sigma_{ZP}^{off}$)}.}
 \begin{center}
\label{tablezps}
\begin{tabular}{lccc}
\hline
\hline
\,\,\,\,\,\, Filter	&	$<ZP_{off}>$	&	$\sigma_{ZP}^{off}$ \\	
\hline
(1) uJAVA	&	0.009  &  0.033    \\	
(2) J0378	&	0.002  &  0.034   \\	
(3) J0395	&	0.000  &  0.030    \\	
(4) J0410	&	0.028  &  0.018    \\	
(5) J0430	&	0.014  &  0.045    \\	
(6) gSDSS	&	-0.017 &  0.020   \\	
(7) J0515	&	-0.009  &  0.010   \\	
(8) rSDSS	&	-0.014 &  0.024    \\	
(9) J0660	&	-0.002  &  0.024    \\	
(10) iSDSS	&	0.002  &  0.033    \\	
(11) J0861	&	0.014  &  0.030    \\	
(12) zSDSS	&	-0.022  &  0.016   \\	
\hline
\hline
\end{tabular}
\end{center}
\end{table}  
 
\subsection{Photometric-noise check.}
\label{noise}
Following the same philosophy as in the previous section, we take advantage of our SED-fitting based photo-z algorithm to check the level of agreement (or disagreement) between the reported photometric noise of sources in our catalogs and the average differences between observed and predicted colours. This exercise serves to understand if there might be additional sources of uncertainties in our photometry, than those already flagged and corrected in the S-PLUS photometric pipeline. 

\vspace{0.2cm}

\begin{figure*}
\includegraphics[width=17.0cm]{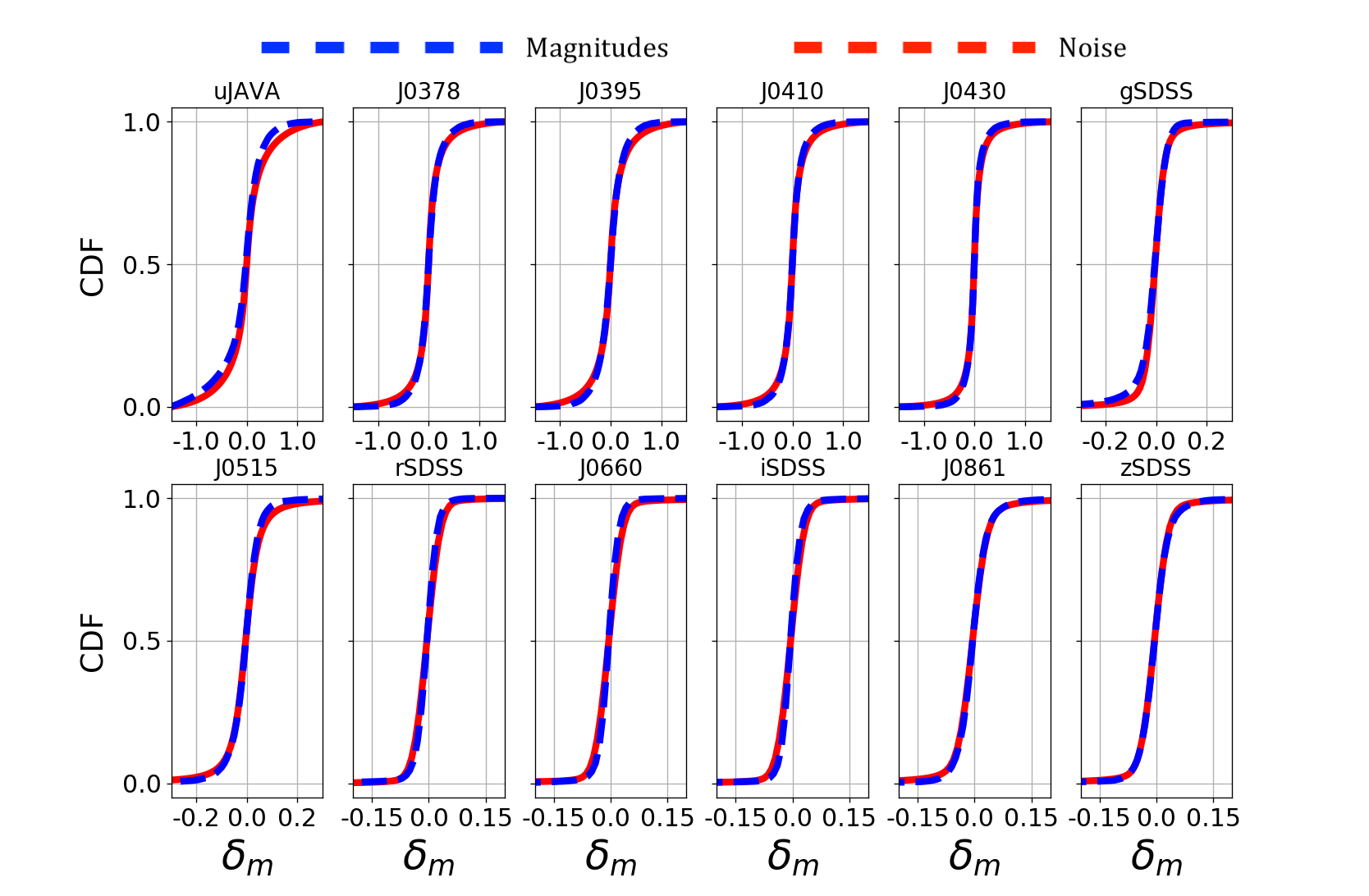}
\caption[Testing photometric noise.]{The figure shows, for every individual filter, the error distribution between the expected and observed colours (i.e., $\delta_{m}$) for the spectroscopic redshift galaxy sample (gray histogram), together with the global photometric error distribution (red dashed line). These distributions correspond to global values averaged over the 90 independent fields and sources with magnitudes 14$<r<$19.}
\label{errordistr}
\end{figure*}

In Figure \ref{errordistr} we show, for every individual filter, the error distribution between the expected and observed colours (i.e., $\delta_{m}$) for the spectroscopic redshift galaxy sample (gray histogram), together with the global photometric error distribution (red dashed line). The latter is calculated as the square root of the quadratic sum of Gaussian functions of width the reported photometric noise of every detection in a given filter. These distributions correspond to global values averaged over the 170 independent fields and sources with magnitudes 14$<r<$19. This analysis serves to demonstrate that, after recalibrating the photometric noise of images, the photometric uncertainties reported for each detection match the observed dispersion between data and models computed during the SED-fitting procedure. Interestingly, we notice that several filters (e.g., $uJava$, $J0515$, $J0861$) show extended tails, sometimes asymmetric, which cannot be explained by regular Poisson noise. This excess signal may arise from additional sources of uncertainties not reported during the image reduction and photometry extraction.

\subsection{Photometric Completeness matrices.}
\label{completABz}

In order to be able to compute the redshift distribution of galaxies in the Stripe-82 region (\textcolor{blue}{Azanha et al., in prep.}), it is necessary to previously derive the expected completeness function of our observations; as much as a function of the magnitude as the redshift. These estimates will serve to compensate the apparent (observed) number counts in our catalogs due to the limited depth of our observations.

\vspace{0.2cm}

As motivated in Section \ref{controlsample}, we can take advantage of the photometric depth of the spectroscopic redshift sample of galaxies in the Stripe-82 to carry out this exercise, since it has a similar photometric depth to that of our observations. Despite the fact the spectroscopic redshift sample utilized in this work is by no means complete in magnitude and redshift (i.e., it does not include all galaxies in the Stripe-82 area), it does contain a large number of galaxies at faint magnitudes and high redshift ranges. Therefore, it may give us a first order characterization of the expected incompleteness at different redshift/magnitude ranges. These completeness functions will be improved in the future once the S-PLUS survey has covered other sky regions with deeper spectroscopic redshift samples such as VVDS \citep{2004A&A...428.1043L} or DEEP2 \citep{2013ApJS..208....5N}. In order to improve the characterization of our selection functions, we have split the spectroscopic redshift sample among early and late-type galaxies, where this classification is based on the most likely spectral-type for each galaxy according to \texttt{BPZ2} code. 

\vspace{0.2cm}

\begin{figure}
\includegraphics[width=7.5cm]{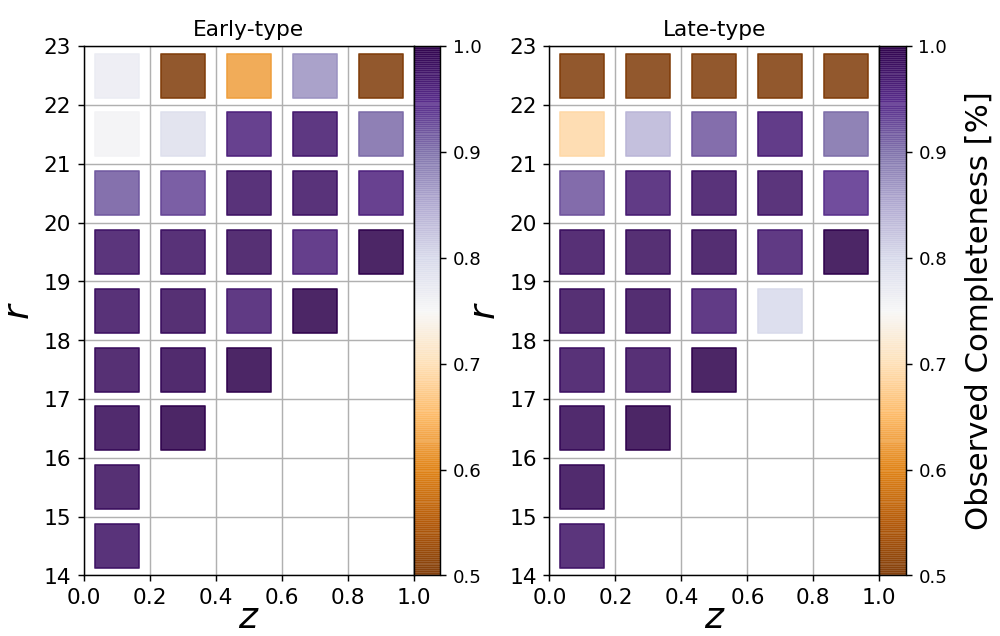}
\caption[Photometric Redshift Completeness]{Photometric Redshift Completeness for our observations. The fraction of detected spectroscopic galaxies, as a function of the magnitude ($r$) and redshift ($z$), is colour-coded as indicated by the vertical colorbar.}
\label{zcompleteness}
\end{figure}

As seen in Figure \ref{zcompleteness}, we define a magnitude-redshift grid where, for each interval, we calculate the fraction of galaxies with spectroscopic redshifts that were also detected in our images. This computation needs to take into account solely common areas, i.e., to differentiate between not detected (i.e., below the detection threshold) from non-observed (i.e., outside the S-PLUS footprint). The so-computed matrix can be converted into a 1-dimensional array where the redshift completeness is computed for sources within a certain magnitude bin. In short, we observe that our observations might be fairly complete up to a redshift $z$ $<$ 0.5 and down to a magnitude $r$ $<$20, for both early- and late-type galaxies. 

\section{Suggestions on how to handle the S-PLUS photo-z}
\label{tips}
 
In order to help users retrieving more reliable samples of galaxies for their science cases, here we include a number of suggestions:

\begin{itemize}
\item Multi-band Photometric catalogs include a Photometric-Quality Flag (based on the \texttt{SExtractor} code), which can be used to exclude objects subject to have a compromised photometry. Sources with very bright nearby objects (i.e., such as saturated stars), might be either removed from the analysis or, in case of necessity, to adopt the smallest apertures to guarantee a high signal-to-noise measurement.

\item The \texttt{BPZ2} code provides a redshift estimate for every source in the catalog, irrespectively of its true nature. In order to decontaminate galaxy samples from galactic sources, we recommend either to adopt a rigorous selection criteria such as the one presented in \textcolor{blue}{Costa-Duarte et al., (in prep.)}, or to treat sources in a flexible statistical framework, where the contribution of each detection to a given analysis is weighted according to the probabilities of being a star or a galaxy. 

\item In order to select a clean and reliable galaxy sample, we encourage the user to use the \texttt{Odds} parameter. Information displayed in Table \ref{phzacctable1}, may help understanding the selection criteria required for a given analysis in terms of photometric redshift error, bias or contamination. 

\item Sources with relatively high $\chi^{2}$ values from the SED-fitting analysis might be treated with care. These sources may represent either objects with faulty photometry or with a SED different from those considered in this work. 

\item Although \texttt{BPZ2} provides the most likely redshift and a confident redshift interval for each galaxy, it is worth stressing that most galaxies show several solutions compatible with the photometric data. Typically, these secondary peaks in redshift correspond to galaxies with different spectral-types. Therefore, we encourage the users not to treat photometric redshifts as simple point-estimates but rather as a multidimensional distribution functions.
\end{itemize} 

\section{Data Access}
\label{dataaccess}

The S-PLUS multi-band photometric catalogs, including photometric redshift estimates, redshift Probability Distribution Functions, Completeness functions and $k$-corrections, along with other additional value-added products, such as stellar-mass content and absolute magnitudes for galaxies and a star/galaxy classification, can be accessed through the following url: \url{https://datalab.noao.edu/splus/}. 

\section{Summary}
\label{conclussions}

In this work, we make use of the Data Release 1 (DR1) and the well-tested Bayesian Photometric Redshift (\texttt{BPZ2}) code to compute and characterize the expected precision of the S-PLUS photometric redshifts. The public photometric catalogue utilized for this exercise corresponds to a nominal area of 336 deg$^{2}$ along the Stripe-82, divided in 170 individual and contiguous pointings, observed with a 5 broad + 7 narrow-band photometric passbands, with a typical photometric depth of $r=21$ (AB) magnitudes. The catalogue (presented in previous papers), includes up to $\sim$3M detected sources, where 1/3 has been classified as potential extragalactic sources. Besides, we take advantage of the abundant spectroscopic redshift information available in this sky region, to compile a large and suitable control sample of $\sim$ 100.000 galaxies with known spectroscopic redshifts, down to a magnitude $r=22$ and up to a redshift $z=1$, to compare our estimates with. 

\vspace{0.2cm}

In order to facilitate their interpretation, the results of this comparison are expressed in terms of basic astronomical variables such as the $r$-band magnitude, the redshift or the spectral-type of galaxies, and the \texttt{BPZ2} \texttt{Odds} parameter. Although there is an in-depth description of our findings in the Appendix section, here we outline several results. We find a photometric redshift precision of $\sigma_{z}\leq0.8\%$ or $\sigma_{z}\leq2.0\%$ for galaxies with magnitudes r$<$17 and r$<$19, respectively. Similarly, a precision of $\sigma_{z}\leq1.5\%$ and $\sigma_{z}\leq3.0\%$ is found for galaxies with a redshift z$<$0.05 and z$<$0.5, respectively. Interestingly, early-type galaxies at z$<$0.1 and z$<$0.5 reach a precision of $\sigma_{z}\leq1.0\%$ and $\sigma_{z}\leq2.0\%$. Likewise, we find a precision of $\sigma_{z}\leq0.8\%$ and $\sigma_{z}\leq1.5\%$ for galaxies with values \texttt{Odds}$>$0.9 and \texttt{Odds}$>$0.6, respectively. The brightest early-type galaxies in our catalogue reach a superb precision of $\sigma_{z}\leq0.6\%$. In all the aforementioned cases, our photo-z estimates show a negligible bias ($\mu_{z}$) and a fraction of catastrophic outliers ($\eta$) inferior to 1$\%$. Based on these results, we forecast a total of $\sim$2M galaxies with a precision $\sigma{z}\leq$0.01 or 16M galaxies with a $\sigma{z}\leq$0.02, in the S-PLUS survey once the entire footprint is observed. 

\vspace{0.2cm}

We identify a redshift window (0.26$<z<$0.32) where our photometric redshift estimates double its precision, due to the simultaneous detection of both [OIII] and H$\alpha$ emission-lines in the $J0660$ \& $J0861$ narrow-band filters. This fact brings a window opportunity in S-PLUS to conduct statistical studies such as luminosity functions, given the powerful combination of both photometric redshift precision and cosmological volume surveyed at that redshift depth. 

\vspace{0.2cm}

In order to fully exploit the information provided by our multi-band photometry, besides these point-like estimates, in this work we also compute the full Probability Distribution Function (i.e., PDF) provided by the \texttt{BPZ2} code for each detection. As demonstrated in this work, these bi-dimensional (i.e., redshift-spectral-type) distributions can successfully encode statistically the redshift uncertainties, and be used to recover the galaxy redshift distribution of galaxies at z $<$ 0.4, with unprecedented precision for a photometric survey in the Southern hemisphere. 

\vspace{0.2cm}

In the final sections of this work, an effort is devoted to double-check the quality of the input photometry, finding no systematic effects such as substantial zero-point corrections or aperture effects. The comparison of expected and observed colours for galaxies with known redshifts indicates that photometric uncertainties in catalogues are properly calibrated. Finally, in order to help deriving statistical analysis in the S-PLUS survey, an effort is made to characterize the photometric completeness of the survey in terms of both $r$-band magnitude and redshift. These and other complementary materials are available online. 

\section{Acknowledgment}
\label{acknowledgment}
 
AMB would like to acknowledge the financial support of the Brazilian funding agency FAPESP (Post-doc fellowship - process number 2014/11806-9), and his colleague Dr. Carlos L\'opez-Sanjuan for all his valuable advices regarding PDF analysis. Likewise, AMB acknowledges Ulisses Manzo and Carlos E. Paladini at IAG/USP for helping with software installation. In addition, AMB would like to dedicate this work to his first son Tiago who was born in Brazil during the time this paper was written. MVCD thanks his scholarship from FAPESP (processes 2014/18632-6 and 2016/05254-9). This work has made use of the computing facilities of the Laboratory of Astroinformatics (IAG/USP, NAT/Unicsul), whose purchase was made possible by the Brazilian agency FAPESP (grant 2009/54006-4) and the INCT-A. The S-PLUS project, including the T80S robotic telescope and the S-PLUS scientific survey, was founded as a partnership between the Fundac\~ao de Amparo a Pesquisa do Estado de S\~ao Paulo (FAPESP), the Observat\'orio Nacional (ON), the Federal University of Sergipe (UFS), and the Federal University of Santa Catarina (UFSC), with important financial and practical contributions from other collaborating institutes in Brazil, Chile (Universidad de La Serena), and Spain (Centro de Estudios de F\'isica del Cosmos de Arag\'on, CEFCA). LSJ acknowledges support from FAPESP and CNPq. JLNC is grateful for financial support received from the GRANT PROGRAMS FA9550-15-1-0167 and FA9550-18-1-0018 of the Southern Office of Aerospace Research and development (SOARD), a branch of the Air Force Office of the Scientific Research International Office of the United States (AFOSR/IO). ADMD thanks FAPESP for financial support. SA acknowledgments financial support from Conselho Nacional de Desenvolvimento Cient\'ifico e Tecnol\'ogico (CNPq, grant 454794/2015-0). AAC acknowledges support from FAPERJ (grant E26/203.186/2016) and CNPq (grants 304971/2016-2 and 401669/2016-5). L.A.A. would like to acknowledge the financial support from FAPESP (N 2012/09716-6, 2013/18245-0). ACS acknowledges funding from CNPq (403580/2016-1, 311153/2018-6) and FAPERGS (17/2551-0001). J.\,A.\,H.\,J. thanks to Brazilian  institution CNPq for financial support through  postdoctoral fellowship (project 150237/2017-0) and Chilean institution CONICYT, Programa de Astronom\'ia, Fondo ALMA-CONICYT 2017, C\'odigo de proyecto 31170038. PAAL thanks the support of CNPq, grant 309398/2018-5.

\bibliographystyle{mnras} 
\bibliography{bibliography}

\vspace{1.0cm}

\noindent$^{1}$ Universidade de S\~{a}o Paulo, IAG, Rua do Matão 1225, S\~{a}o Paulo, SP, Brazil \\
$^{2}$ Departamento de Astronomia, Instituto de F\'isica, Universidade Federal do Rio Grande do Sul (UFRGS), Av. Bento Gon\c{c}alves 9500, \\
$^{3}$ Instituto de F\'isica, Universidade de S\~ao Paulo, Rua do Mat\~ao 1371, S\~ao Paulo, SP 05508-090, Brazil \\
$^{4}$ Departamento de F\'isica y Astronom\'ia, Facultad de Ciencias, Universidad de La Serena, Av. Juan Cisternas 1200 Norte, La Serena, Chile. \\ 
$^{5}$ Instituto de Investigaci\'on Multidisciplinario en Ciencia y Tecnolog\'ia, Universidad de La Serena. Avenida Juan Cisternas 1400, La Serena, Chile \\
$^{6}$ NOAO, P.O. Box 26732, Tucson, AZ 85726 \\
$^{7}$ Departamento de F\'isica, Universidade Federal de Santa Catarina, Florian\'{o}polis, SC, 88040-900, Brazil \\
$^{8}$ Departamento de Ciencias F\'isicas, Universidad Andr\'es Bello, Fern\'andez Concha 700, Las Condes, Santiago, Chile \\
$^{9}$ Valongo Observatory, Federal University of Rio de Janeiro, Ladeira Pedro Antonio 43, Saude Rio de Janeiro, RJ, 20080-090, Brazil \\
$^{10}$ Instituto de Matem\'{a}tica, Estat\'{i}stica e F\'{i}sica, Universidade Federal do Rio Grande, Rio Grande 96203-900, Brazil\\ 
$^{11}$ Observat\'orio Nacional/MCTIC, Rua Gen. Jos\'{e} Cristino, 77, 20921-400, Rio de Janeiro, Brazil \\
$^{12}$ Departamento de F\'isica, Universidade Federal de Sergipe, Av. Marechal Rondon, S/N, 49000-000 S\~ao Crist\'ov\~ao, SE, Brazil \\
$^{13}$ Center for Space Science and Technology, University of Maryland, Baltimore County, 1000 Hilltop Circle, Baltimore, MD 21250, USA \\
$^{14}$ Center for Earth System Science, National Institute for Space Research, Av. dos Astronautas, 1758, S\~ao Jos\'e dos Campos, Brazil \\
$^{15}$ Departamento de F\'isica Matem\'atica, Instituto de F\'isica, Universidade de S\~ao Paulo, Rua do Mat\~ao 1371, CEP 05508-090  \\
$^{16}$ Instituto de Astrof\'isica de Andaluc\'ia. IAA-CSIC. Glorieta de la astronom\'ia S/N. 18008, Granada, Spain. \\
$^{17}$ Departamento de F\'isica Te\'orica e Experimental, Universidade Federal do Rio Grande do Norte, CP 1641, Natal, RN, 59072-970, Brazil  \\
$^{18}$ Department of Physics \& Astronomy, University of North Carolina at Chapel Hill, NC 27599-3255, USA \\
$^{19}$ Universidade Federal do Paran\'a, Campus Jandaia do Sul, Rua Dr. Jo\~ao Maximiano, 426, Jandaia do Sul-PR, 86900-000, Brazil \\
$^{20}$ Centro Brasileiro de Pesquisas F\'isicas, Rua Dr. Xavier Sigaud 150, Rio de Janeiro, RJ, CEP 22290-180, Brazil \\
$^{21}$ Department of Physics, University of Notre Dame, Notre Dame, IN 46556, USA \\ 
$^{22}$ JINA Center for the Evolution of the Elements (JINA-CEE), USA \\

\appendix

\section{Photometric Redshift Performance: Tables}
\label{photozTables}

In Table \ref{phzacctable0} we present a general overview of the photometric redshift performance as a function of the magnitude, redshift and \texttt{Odds}. In Table \ref{phzacctable1}, \ref{phzacctable2} and \ref{phzacctable3} this information is extended defining thinner bins in magnitude, redshift and \texttt{Odds}, respectively, Finally, we present in Table \ref{bpzkcorrections} several $K$-corrections computed with the \texttt{BPZ2} templates in the $uJAVA$ \& the $rSDSS$ filters.

\begin{table*}
\caption{S-PLUS General Table: The table shows the bias ($\mu_{z}$), the precision ($\sigma_{z}$), and the fraction of outliers ($\eta_{z}$) for all types of galaxies (i.e., $all$), early (i.e., $red$) and late (i.e., $blue$) spectral-types and the fraction of galaxies (\#) as a function of the $r$-band magnitude, the redshift and the \texttt{Odds} parameter.}
\begin{center}
\label{phzacctable0}
\begin{tabular}{|c|c|c|c|c|c|c|c|c|c|c|c|c|}
\hline
\hline
\texttt{r}, \texttt{z}, \texttt{Odds}  & $\mu_{z}^{all}$ & $\sigma_{z}^{all}$ & $\eta_{z}^{all}$ & $\#^{all}$ & $\mu_{z}^{red}$ & $\sigma_{z}^{red}$ & $\eta_{z}^{red}$ & $\#^{red}$ & $\mu_{z}^{blue}$ & $\sigma_{z}^{blue}$ & $\eta_{z}^{blue}$ & $\#^{blue}$ \\
\hline
 r $<$ 16.0 & 0.002 & 0.009 & 0.004 & 1  &  0.001 & 0.006 & 0.002 & 0 &  0.003 & 0.013 & 0.007  & 0 \\ 
 r $<$ 17.0 & 0.001 & 0.011 & 0.007 & 6  &  -0.000 & 0.009 & 0.001 & 2 &  0.003 & 0.013 & 0.013 & 3 \\ 
 r $<$ 18.0 & 0.001 & 0.015 & 0.008 & 19 &  -0.001 & 0.012 & 0.002 & 8 &  0.003 & 0.017 & 0.013 & 10 \\ 
 r $<$ 19.0 & 0.002 & 0.020 & 0.014 & 41 &  -0.001 & 0.014 & 0.003 & 15 &  0.004 & 0.024 & 0.020 & 25 \\ 
 r $<$ 20.0 & 0.002 & 0.026 & 0.025 & 69 &  -0.002 & 0.019 & 0.006 & 25 &  0.005 & 0.031 & 0.037 & 44 \\ 
 r $<$ 21.0 & 0.000 & 0.030 & 0.036 & 88 &  -0.004 & 0.023 & 0.010 & 35 &  0.004 & 0.035 & 0.053 & 53 \\ 
\hline
\hline
 $z<$ 0.05 & 0.015 & 0.027 & 0.063 & 5 &  0.003 & 0.010 & 0.005 & 1 &  0.025 & 0.040 & 0.080 & 3 \\
 $z<$ 0.10 & 0.010 & 0.024 & 0.039 & 17 &  0.002 & 0.011 & 0.002 & 4 &  0.016 & 0.031 & 0.051 & 13 \\
 $z<$ 0.20 & 0.006 & 0.025 & 0.031 & 44 &  0.000 & 0.015 & 0.002 & 12 &  0.010 & 0.030 & 0.042 & 32 \\ 
 $z<$ 0.30 & 0.004 & 0.026 & 0.031 & 58 &  -0.000 & 0.016 & 0.004 & 17 &  0.007 & 0.031 & 0.042 & 41  \\ 
 $z<$ 0.40 & 0.003 & 0.026 & 0.029 & 69 &  -0.001 & 0.018 & 0.005 & 23 &  0.007 & 0.032 & 0.041 & 46 \\ 
 $z<$ 0.50 & 0.002 & 0.028 & 0.028 & 78 &  -0.002 & 0.021 & 0.006 & 29 &  0.005 & 0.033 & 0.041 & 49 \\ 
 $z<$ 1.00 & 0.000 & 0.030 & 0.036 & 88 &  -0.004 & 0.023 & 0.010 & 35 &  0.004 & 0.035 & 0.053 & 53 \\ 
\hline
\hline
\texttt{Odds} $>$ 0.00 & 0.000 & 0.030 & 0.036 & 88 &  -0.004 & 0.023 & 0.010 & 35 &  0.004 & 0.035 & 0.054 & 53  \\ 
\texttt{Odds} $>$ 0.10 & 0.000 & 0.029 & 0.034 & 86 &  -0.004 & 0.023 & 0.010 & 35 &  0.004 & 0.034 & 0.050 & 51 \\ 
\texttt{Odds} $>$ 0.20 & 0.000 & 0.026 & 0.027 & 80 &  -0.004 & 0.022 & 0.007 & 33 &  0.004 & 0.030 & 0.041 & 46 \\ 
\texttt{Odds} $>$ 0.30 & 0.001 & 0.023 & 0.023 & 70 &  -0.003 & 0.019 & 0.005 & 29 &  0.004 & 0.027 & 0.036 & 40 \\ 
\texttt{Odds} $>$ 0.40 & 0.001 & 0.020 & 0.021 & 57 &  -0.002 & 0.016 & 0.004 & 24 &  0.004 & 0.024 & 0.034 & 33 \\ 
\texttt{Odds} $>$ 0.50 & 0.001 & 0.017 & 0.020 & 44 &  -0.001 & 0.014 & 0.004 & 17 &  0.003 & 0.020 & 0.032 & 26 \\ 
\texttt{Odds} $>$ 0.60 & 0.001 & 0.015 & 0.019 & 32 &  -0.001 & 0.012 & 0.003 & 12 &  0.003 & 0.017 & 0.029 & 19 \\ 
\texttt{Odds} $>$ 0.70 & 0.001 & 0.013 & 0.019 & 23 &  -0.001 & 0.010 & 0.003 & 9 &  0.002 & 0.015 & 0.029 & 14  \\ 
\texttt{Odds} $>$ 0.80 & 0.001 & 0.011 & 0.019 & 16 &  -0.000 & 0.009 & 0.003 & 6 &  0.002 & 0.013 & 0.031 & 9 \\ 
\texttt{Odds} $>$ 0.90 & 0.001 & 0.009 & 0.020 & 9 &  0.000 & 0.008 & 0.002 & 3 &  0.002 & 0.010 & 0.032 & 5  \\ 
\texttt{Odds} $>$ 0.95 & 0.001 & 0.008 & 0.017 & 5 &  0.001 & 0.007 & 0.001 & 2 &  0.002 & 0.009 & 0.027 & 3 \\ 
\texttt{Odds} $>$ 0.99 & 0.001 & 0.007 & 0.009 & 2 &  0.001 & 0.006 & 0.000 & 0 &  0.002 & 0.008 & 0.016 & 1 \\
\hline
\hline
\end{tabular}
\end{center}
\end{table*}

\begin{table*}
\caption{The table shows the bias ($\mu_{z}$), the precision ($\sigma_{z}$), and the fraction of outliers ($\eta_{z}$) for all types of galaxies (i.e., $all$), early (i.e., $red$) and late (i.e., $blue$) spectral-types and the fraction of galaxies (\#) for different magnitude intervals.}
\begin{center}
\label{phzacctable1}
\begin{tabular}{|c|c|c|c|c|c|c|c|c|c|c|c|c|}
\hline
\hline
\texttt{r} & $\mu_{z}^{all}$ & $\sigma_{z}^{all}$ & $\eta_{z}^{all}$ & $\#^{all}$ & $\mu_{z}^{red}$ & $\sigma_{z}^{red}$ & $\eta_{z}^{red}$ & $\#^{red}$ & $\mu_{z}^{blue}$ & $\sigma_{z}^{blue}$ & $\eta_{z}^{blue}$ & $\#^{blue}$ \\
\hline
 14.5 $<$ $\texttt{r}$ $<$ 15.5 & 0.001 & 0.008 & 0.005 & 0 &  0.000 & 0.006 & 0.003 & 0 &  0.003 & 0.010 & 0.006 & 0 \\
 15.5 $<$ $\texttt{r}$ $<$ 16.5 & 0.002 & 0.010 & 0.007 & 2 &  0.000 & 0.008 & 0.001 & 1 &  0.003 & 0.012 & 0.012 & 1 \\  
 16.5 $<$ $\texttt{r}$ $<$ 17.5 & 0.001 & 0.014 & 0.009 & 7 &  -0.001 & 0.012 & 0.001 & 3 &  0.003 & 0.016 & 0.015 & 4 \\  
 17.5 $<$ $\texttt{r}$ $<$ 18.5 & 0.001 & 0.020 & 0.010 & 18 &  -0.002 & 0.016 & 0.003 & 7 &  0.004 & 0.024 & 0.015 & 11 \\  
 18.5 $<$ $\texttt{r}$ $<$ 19.5 & 0.003 & 0.032 & 0.029 & 26 &  -0.003 & 0.025 & 0.007 & 7 &  0.007 & 0.036 & 0.039 & 18 \\  
 19.5 $<$ $\texttt{r}$ $<$ 20.5 & -0.003 & 0.045 & 0.064 & 23 &  -0.011 & 0.034 & 0.018 & 9 &  0.004 & 0.054 & 0.093 & 14 \\  
 20.5 $<$ $\texttt{r}$ $<$ 21.5 & -0.018 & 0.047 & 0.061 & 16 &  -0.017 & 0.038 & 0.019 & 10 &  -0.022 & 0.076 & 0.143 & 5 \\  
\hline
\hline
\end{tabular}
\end{center}
\end{table*}

\begin{table*}
\caption{The table shows the bias ($\mu_{z}$), the precision ($\sigma_{z}$), and the fraction of outliers ($\eta_{z}$) for all types of galaxies (i.e., $all$), early (i.e., $red$) and late (i.e., $blue$) spectral-types and the fraction of galaxies (\#) for different redshift intervals.}
\begin{center}
\label{phzacctable2}
\begin{tabular}{|c|c|c|c|c|c|c|c|c|c|c|c|c|}
\hline
\hline
\texttt{z} & $\mu_{z}^{all}$ & $\sigma_{z}^{all}$ & $\eta_{z}^{all}$ & $\#^{all}$ & $\mu_{z}^{red}$ & $\sigma_{z}^{red}$ & $\eta_{z}^{red}$ & $\#^{red}$ & $\mu_{z}^{blue}$ & $\sigma_{z}^{blue}$ & $\eta_{z}^{blue}$ & $\#^{blue}$ \\
\hline
0.0 $<$ $z$ $<$ 0.1 & 0.010 & 0.024 & 0.041 & 17 &  0.003 & 0.012 & 0.004 & 4 &  0.016 & 0.031 & 0.052 & 13 \\  
0.1 $<$ $z$ $<$ 0.2 & 0.003 & 0.026 & 0.026 & 27 &  -0.002 & 0.016 & 0.003 & 8 &  0.007 & 0.031 & 0.036 & 18 \\ 
0.2 $<$ $z$ $<$ 0.3 & -0.002 & 0.028 & 0.029 & 13 &  -0.002 & 0.020 & 0.008 & 4 &  -0.002 & 0.033 & 0.040 & 8 \\ 
0.3 $<$ $z$ $<$ 0.4 & -0.002 & 0.029 & 0.019 & 10 &  -0.003 & 0.025 & 0.007 & 6 &  -0.000 & 0.037 & 0.034 & 4 \\ 
0.4 $<$ $z$ $<$ 0.5 & -0.017 & 0.035 & 0.023 & 9 &  -0.016 & 0.031 & 0.010 & 5 &  -0.022 & 0.044 & 0.046 & 3 \\ 
0.5 $<$ $z$ $<$ 0.8 & -0.032 & 0.050 & 0.091 & 9 &  -0.023 & 0.036 & 0.026 & 6 &  -0.075 & 0.089 & 0.209 & 3 \\ 
0.8 $<$ $z$ $<$ 1.0 & -0.206 & 0.103 & 0.674 & 0 &  -0.199 & 0.075 & 0.694 & 0 &  -0.223 & 0.140 & 0.661 & 0 \\ 
\hline
\hline
\end{tabular}
\end{center}
\end{table*}

\begin{table*}
\caption{The table shows the bias ($\mu_{z}$), the precision ($\sigma_{z}$), and the fraction of outliers ($\eta_{z}$) for all types of galaxies (i.e., $all$), early (i.e., $red$) and late (i.e., $blue$) spectral-types and the fraction of galaxies (\#) for different \texttt{Odds} intervals.}
\begin{center}
\label{phzacctable3}
\begin{tabular}{|c|c|c|c|c|c|c|c|c|c|c|c|c|}
\hline
\hline
\texttt{Odds} & $\mu_{z}^{all}$ & $\sigma_{z}^{all}$ & $\eta_{z}^{all}$ & $\#^{all}$ & $\mu_{z}^{red}$ & $\sigma_{z}^{red}$ & $\eta_{z}^{red}$ & $\#^{red}$ & $\mu_{z}^{blue}$ & $\sigma_{z}^{blue}$ & $\eta_{z}^{blue}$ & $\#^{blue}$ \\
\hline
0.0 $<$ $\texttt{Odds}$ $<$ 0.1 & -0.021 & 0.105 & 0.159 & 1 &  -0.004 & 0.023 & 0.010 & 35 &  0.004 & 0.035 & 0.054 & 53 \\ 
0.1 $<$ $\texttt{Odds}$ $<$ 0.2 & -0.019 & 0.087 & 0.115 & 6 &  -0.004 & 0.023 & 0.010 & 35 &  0.004 & 0.034 & 0.050 & 51 \\ 
0.2 $<$ $\texttt{Odds}$ $<$ 0.3 & -0.008 & 0.058 & 0.055 & 10 &  -0.004 & 0.022 & 0.007 & 33 &  0.004 & 0.030 & 0.041 & 46 \\ 
0.3 $<$ $\texttt{Odds}$ $<$ 0.4 & -0.003 & 0.042 & 0.031 & 12 &  -0.003 & 0.019 & 0.005 & 29 &  0.004 & 0.027 & 0.036 & 40 \\ 
0.4 $<$ $\texttt{Odds}$ $<$ 0.5 & 0.000 & 0.033 & 0.025 & 13 &  -0.002 & 0.016 & 0.004 & 24 &  0.004 & 0.024 & 0.034 & 33 \\ 
0.5 $<$ $\texttt{Odds}$ $<$ 0.6 & 0.001 & 0.028 & 0.025 & 11 &  -0.001 & 0.014 & 0.004 & 17 &  0.003 & 0.020 & 0.032 & 26 \\ 
0.6 $<$ $\texttt{Odds}$ $<$ 0.7 & 0.001 & 0.022 & 0.019 & 8 &  -0.001 & 0.012 & 0.003 & 12 &  0.003 & 0.017 & 0.029 & 19 \\ 
0.7 $<$ $\texttt{Odds}$ $<$ 0.8 & 0.001 & 0.018 & 0.017 & 7 &  -0.001 & 0.010 & 0.003 & 9 &  0.002 & 0.015 & 0.029 & 14 \\ 
0.8 $<$ $\texttt{Odds}$ $<$ 0.9 & 0.001 & 0.014 & 0.019 & 6 &  -0.000 & 0.009 & 0.003 & 6 &  0.002 & 0.013 & 0.031 & 9 \\ 
0.9 $<$ $\texttt{Odds}$ $<$ 1.0 & 0.001 & 0.009 & 0.020 & 9 &  0.000 & 0.008 & 0.002 & 3 &  0.002 & 0.010 & 0.032 & 5 \\ 
\hline
\hline
\end{tabular}
\end{center}
\end{table*}

\begin{table*}
\caption{$K$-corrections for the \texttt{BPZ2} templates utilized in this work, for the $uJAVA$ \& the $rSDSS$ filters. An on-line version of this table (including all filters) is available at the following link: \url{https://datalab.noao.edu/splus/}}
\begin{center}
\label{bpzkcorrections}
\begin{tabular}{|c|c|c|c|c|c|c|c|c|c|c|c|c|c|c|}
\hline
\hline
$uJAVA$  & T$_{1}$ & T$_{2}$ & T$_{3}$ & T$_{4}$ & T$_{5}$ & T$_{6}$ & T$_{7}$ & T$_{8}$ & T$_{9}$ & T$_{10}$ & T$_{11}$ & T$_{12}$ & T$_{13}$ & T$_{14}$ \\
\hline
0.0 $<$ $z$ $<$ 0.1 & 0.15 & 0.16 & 0.19 & 0.19 & 0.21 & 0.19 & 0.17 & 0.16 & 0.15 & 0.11 & 0.12 & 0.13 & 0.10 & 0.07 \\
0.1 $<$ $z$ $<$ 0.2 & 0.50 & 0.55 & 0.58 & 0.63 & 0.69 & 0.64 & 0.54 & 0.50 & 0.46 & 0.33 & 0.32 & 0.19 & 0.12 & 0.00 \\
0.2 $<$ $z$ $<$ 0.3 & 1.04 & 1.08 & 1.11 & 1.20 & 1.27 & 1.16 & 1.00 & 0.91 & 0.86 & 0.68 & 0.47 & 0.24 & 0.13 & 0.00 \\
0.3 $<$ $z$ $<$ 0.4 & 1.63 & 1.69 & 1.73 & 1.86 & 1.93 & 1.64 & 1.39 & 1.26 & 1.18 & 0.93 & 0.66 & 0.30 & 0.15 & -0.01 \\
0.4 $<$ $z$ $<$ 0.5 & 2.13 & 2.25 & 2.32 & 2.52 & 2.56 & 1.97 & 1.65 & 1.49 & 1.38 & 1.08 & 0.88 & 0.41 & 0.14 & -0.06 \\
0.5 $<$ $z$ $<$ 0.8 & 2.29 & 2.47 & 2.58 & 2.83 & 2.83 & 1.89 & 1.52 & 1.36 & 1.25 & 0.95 & 0.91 & 0.48 & 0.13 & -0.15 \\
0.8 $<$ $z$ $<$ 1.0 & 3.19 & 3.52 & 3.73 & 4.00 & 3.56 & 1.95 & 1.54 & 1.36 & 1.24 & 0.91 & 0.99 & 0.30 & 0.00 & -0.29 \\
1.0 $<$ $z$ $<$ 1.2 & 3.95 & 4.43 & 4.73 & 4.80 & 3.82 & 1.94 & 1.54 & 1.35 & 1.22 & 0.87 & 1.22 & 0.31 & -0.01 & -0.34 \\
%\hline
\hline
$rSDSS$  & T$_{1}$ & T$_{2}$ & T$_{3}$ & T$_{4}$ & T$_{5}$ & T$_{6}$ & T$_{7}$ & T$_{8}$ & T$_{9}$ & T$_{10}$ & T$_{11}$ & T$_{12}$ & T$_{13}$ & T$_{14}$ \\
\hline
0.0 $<$ $z$ $<$ 0.1 & 0.03 & 0.03 & 0.03 & 0.04 & 0.04 & 0.05 & 0.04 & 0.04 & 0.05 & 0.04 & 0.05 & 0.03 & 0.01 & 0.01 \\ 
0.1 $<$ $z$ $<$ 0.2 & 0.10 & 0.12 & 0.12 & 0.14 & 0.16 & 0.19 & 0.17 & 0.17 & 0.17 & 0.14 & 0.16 & 0.08 & 0.01 & -0.04 \\
0.2 $<$ $z$ $<$ 0.3 & 0.20 & 0.23 & 0.24 & 0.28 & 0.32 & 0.36 & 0.32 & 0.32 & 0.32 & 0.27 & 0.26 & 0.12 & 0.01 & -0.09 \\
0.3 $<$ $z$ $<$ 0.4 & 0.33 & 0.39 & 0.41 & 0.47 & 0.53 & 0.57 & 0.51 & 0.50 & 0.48 & 0.43 & 0.37 & 0.15 & -0.01 & -0.13 \\
0.4 $<$ $z$ $<$ 0.5 & 0.52 & 0.60 & 0.64 & 0.72 & 0.79 & 0.83 & 0.75 & 0.72 & 0.68 & 0.62 & 0.51 & 0.24 & 0.06 & -0.07 \\
0.5 $<$ $z$ $<$ 0.8 & 1.07 & 1.20 & 1.28 & 1.38 & 1.48 & 1.57 & 1.48 & 1.39 & 1.28 & 1.21 & 0.95 & 0.53 & 0.27 & 0.06 \\
0.8 $<$ $z$ $<$ 1.0 & 1.68 & 1.82 & 1.93 & 2.03 & 2.16 & 2.25 & 2.17 & 2.04 & 1.89 & 1.76 & 1.46 & 0.95 & 0.62 & 0.29 \\
1.0 $<$ $z$ $<$ 1.2 & 2.12 & 2.28 & 2.41 & 2.54 & 2.71 & 2.75 & 2.60 & 2.43 & 2.26 & 2.04 & 1.68 & 1.01 & 0.63 & 0.24 \\
\hline
\hline
\end{tabular}
\end{center}
\end{table*}

\end{document}